\documentclass[11pt]{article}
\pdfoutput=1
\usepackage{amsmath,amssymb,amsfonts,amsthm,graphicx, epsf,dsfont,dcolumn,yfonts}
\usepackage{latexsym}
\usepackage[colorlinks=true, pdfstartview=FitV, linkcolor=black, citecolor=black, urlcolor=black]{hyperref}

\newlength{\textlength}
\newlength{\overlinelength}

\newcommand{\be}{\begin{equation}}
\newcommand{\ee}{\end{equation}}
\newcommand{\bea}{\begin{eqnarray}}
\newcommand{\eea}{\end{eqnarray}}
\newcommand{\beq}{\begin{equation}}
\newcommand{\eeq}{\end{equation}}
\newcommand{\beqn}{\begin{eqnarray}}
\newcommand{\eeqn}{\end{eqnarray}}

\def\det{{\rm{det}}}

\def\ads{{\rm AdS}}

\def\G{\Gamma}

\newcommand{\wn}{\textswab{w}\kern1pt}
\newcommand{\qn}{\textswab{q}\kern1pt}

\numberwithin{equation}{section}

\begin{document}

\begin{center}\ \\
\vspace{60pt}

{\Large {\bf Shear Modes, Criticality and Extremal Black Holes}}\\ 

\vspace{50pt}

{\large Mohammad Edalati, Juan I. Jottar and Robert G. Leigh}
\vspace{9pt}

{\it Department of Physics, University of Illinois at Urbana-Champaign,\\ Urbana IL 61801, USA}
\\{\tt edalati,jjottar2,rgleigh@illinois.edu}\\ [4mm]

\end{center}
\vspace{70pt}

\centerline{\bf Abstract}

\noindent We consider a $(2+1)$-dimensional field theory, assumed to be holographically dual to the extremal Reissner-Nordstr\"{o}m AdS$_4$ black hole background, and calculate the retarded correlators of charge (vector) current and energy-momentum (tensor) operators at finite momentum and frequency. We show that, similar to what was observed previously for the correlators of scalar and spinor operators, these correlators exhibit emergent scaling behavior at low frequency. We numerically compute the electromagnetic and gravitational quasinormal frequencies (in the shear channel) of the extremal Reissner-Nordstr\"{o}m AdS$_4$ black hole corresponding to the spectrum of poles in the retarded correlators. The picture that emerges is quite simple: there is a branch cut along the negative imaginary frequency axis, and a series of isolated poles corresponding to damped excitations. All of these poles are always in the lower half complex frequency plane, indicating stability. We show that this analytic structure can be understood as the proper limit of finite temperature results as $T$ is taken to zero holding the chemical potential $\mu$ fixed.

\newpage

\section{Introduction}
The AdS/CFT correspondence is emerging as a promising approach for analyzing strongly-coupled condensed matter systems.\footnote{See \cite{Aharony:1999ti} for a review of the AdS/CFT correspondence.} Although the systems accessible through the duality need a lot of refinement in order to resemble actual condensed matter systems, (more than) a decade-long experience with the correspondence has taught us that it is extremely useful in extracting some universal features of strongly-interacting systems which would be hard, or even impossible, to obtain using field theoretic methods. It is in this sense that applying the tools of the correspondence to study strongly-interacting condensed matter systems seems to be promising. Recent examples of such applications  include modeling superconductivity and superfluidity \cite{Gubser:2008px, Hartnoll:2008vx,Hartnoll:2008kx, Herzog:2008he, Basu:2008st}, engineering systems with Schr\"{o}dinger and Lifshitz symmetries \cite{Son:2008ye, Balasubramanian:2008dm, Kachru:2008yh, Herzog:2008wg, Maldacena:2008wh, Adams:2008wt, Imeroni:2009cs, Adams:2009dm}, or modeling a system which can exhibit non-Fermi liquid type behavior \cite{Lee:2008xf, Liu:2009dm, Faulkner:2009wj, Rey:2008zz, Hartnoll:2009ns}. See \cite{Hartnoll:2009sz, Herzog:2009xv, McGreevy:2009xe} for reviews. 

Using the correspondence to study quantum critical regions of condensed matter systems  is another example of such applications \cite{Herzog:2007ij, Hartnoll:2007ih, Hartnoll:2007ip}. At the quantum critical point, the system is often a strongly-coupled conformal field theory, and it is usually the case that a proper field theoretic understanding is not known. The authors of \cite{Faulkner:2009wj} considered a simple gravitational background whose dual field theory exhibits a variety of emergent quantum critical behaviors. They considered the background of extremal Reissner-Nordstr\"{o}m AdS$_{d+1}$ black hole whose dual field theory is supposed to be a $d$-dimensional strongly-coupled field theory which is at zero temperature and finite U(1) charge density. Considering probe scalars as well as spinors in the geometry, it was shown that the retarded correlator of the dual operators show  emergent quantum criticality at low frequency. The existence of such behaviors was attributed to the fact that the near horizon region of the background geometry (which encodes the IR physics of the boundary theory) is ${\rm AdS}_2\times\mathds{R}^2$, and the assumption that there exists some sort of IR CFT dual to the AdS$_2$ region. It was shown that the behavior of the retarded Green function $G_R(\omega,\vec k)$ at low frequency is dictated by $G_R(\omega=0,\vec k)$ as well as the conformal dimension of some operators in the IR CFT. More specifically, it was shown that $G_R(\omega,\vec k)$ can in general exhibit  a scaling behavior for the spectral density, a log-periodic behavior, or in the case of fermionic operators indicate the existence of Fermi-like surfaces with quasi-particle excitations of non-Fermi liquid type. 

In \cite{Edalati:2009bi} we examined the role played by the IR CFT in constraining the low frequency behavior of retarded correlators of charge vector and energy momentum tensor operators of the boundary theory. The main focus though was on extracting some transport coefficients, such as conductivity and viscosity, of the boundary theory, and investigating the relationship between the IR CFT and the universality of such coefficients.  As such, the analyses were limited to the case of zero spatial momentum. The boundary theory considered was a $(2+1)$-dimensional field theory at zero temperature and finite charge density dual to an extremal Reissner-Nordstrom AdS$_{4}$ black hole. Although temperature is zero, there is a scale in the problem set by the chemical potential for the charge density and moreover the entropy density is finite.\footnote{This is presumably a ``large $N$'' effect. Analysis of this has recently appeared in \cite{Hartnoll:2009ns}.} It was shown in \cite{Edalati:2009bi} that, for example, the ratio of shear viscosity to the entropy density is $1/4\pi$ saturating the KSS bound \cite{Kovtun:2004de}. There are a variety of reasons why such analyses at zero-temperature are interesting. First, calculating transport coefficients  involves taking the limit of small frequency, which at zero temperature becomes subtle and extra care is needed to deal with this subtlety. Secondly, they generalize the universality arguments, given for finite temperature backgrounds \cite{Benincasa:2006fu, Iqbal:2008by, Kovtun:2008kx},  to backgrounds of extremal black holes. Third, the system studied may be in the universality class of some quantum critical points, hence understanding the universal behavior of the transport coefficients of this system may shed light on the physics of quantum critical points. 

In this paper, we analyze the retarded correlators of the charge vector current and energy-momentum (tensor) operators of the aforementioned $(2+1)$-dimensional boundary theory for non-vanishing (spatial) momentum. Although we perform the analyses in the $(2+1)$-dimensional boundary theory, they can easily be carried out for higher dimensional boundary theories dual to extremal Reissner-Nordstrom AdS$_{d+1}$ black hole. Our focus here is on the retarded correlators of those operators which are in the so-called shear channel of the boundary theory. We find that for generic momentum, similar to the cases of scalar and spinor operators studied in \cite{Faulkner:2009wj}, the retarded correlators of such vector and tensor operators exhibit emergent scaling behavior at low frequency. These operators give rise in the IR to two sets of operators with different conformal dimensions. The low frequency scaling behaviors of the boundary theory retarded correlators, especially their scaling exponents,  depend only on one of the two IR CFT operators. By taking linear combinations of the dual bulk modes, we construct the gauge invariant modes appropriate for extracting the quasinormal modes. 

We study the small frequency regime analytically for finite momentum, and we explore the analytic structure of the shear channel retarded Green functions numerically for generic frequency and momentum. We present evidence for the existence of a branch cut in these Green functions at the origin, as well as a series of metastable modes corresponding to isolated poles in the lower half complex frequency plane. By turning on temperature we observe how the branch cut dissolves into a series of  poles on the negative imaginary axis and determine the corresponding dispersion constant of the leading pole, in agreement with previous results. 

The organization of the paper is as follows. In section \ref{sectiontwo}, we review the Reissner-Nordstr\"{o}m AdS$_4$  black hole background, its dual boundary field theory, near horizon ${\rm AdS}_2\times\mathds{R}^2$ geometry and the dual IR CFT.  In section \ref{sectionthree} we present the Linearized Einstein-Maxwell equations for the bulk modes dual to the (vector and tensor) operators in the shear channel of the boundary theory. Taking linear combinations of the modes, we construct the gauge-invariant combinations in terms of which the linearized Einstein-Maxwell equations reduce to a set of two coupled second-order differential equations. By introducing ``master variables" we then decouple these equations. In section \ref{sectionfour} we solve the decoupled equations by matching the solutions in the outer region to the solutions in the inner ${\rm AdS}_2$ region. Having determined the solutions, we calculate the retarded correlators of the charge vector current and energy-momentum tensor operators of the boundary theory. For generic momentum, we then extract the low frequency emergent scaling behavior of their spectral functions. In section \ref{sectionfive} we argue that the (shear-type) electromagnetic and gravitational perturbations of the extremal Reissner-Nordstr\"{o}m AdS$_4$ do not cause instability by showing that the associated quasinormal frequencies are all located in the lower half of the complex frequency plane.  We then numerically compute the spectrum of the above-mentioned quasinormal frequencies and compare the numerical results with the analogous quasinormal frequencies of the non-extremal Reissner-Nordstr\"{o}m AdS$_{4}$ black hole background.  In Appendix A we have summarized the finite-temperature equations we occasionally used in the bulk of the paper in order to extract the shear-type quasinormal frequencies of the non-extremal Reissner-Nordstr\"{o}m AdS$_4$  black hole.

\section{The Background and the Boundary Field Theory}\label{sectiontwo}

Consider the Einstein-Maxwell action in 3+1 spacetime dimensions with a negative cosmological constant $\Lambda = -3/L^2$ 
\bea\label{EMaction}
S= \frac{1}{2\kappa_4^2}\int d^4x \sqrt{-g}\left(R-2\Lambda- L^2F_{\mu\nu}F^{\mu\nu}\right),
\eea
where $L$ is the curvature radius of AdS$_4$. The background we consider is the Reissner-Nordstr\"{o}m AdS$_4$ black hole
\begin{align}\label{metric}
ds^2&=g_{\mu\nu}dx^\mu dx^\nu=\frac{r^2}{L^2}\Big(- f(r) dt^2 + dx^2 + dy^2\Big)+\frac{L^2}{r^2 f(r)} dr^2,\\
A&=\mu\Big(1-\frac{r_0}{r}\Big)dt,\label{gfield} 
\end{align}
which is a solution to the Einstein-Maxwell equations obtained from \eqref{EMaction}, where 
\bea\label{fmu}
f(r)=1-M\left(\frac{r_0}{r}\right)^3+Q^2\left(\frac{r_0}{r}\right)^4, \qquad \qquad \mu=\frac{Qr_0}{L^2}.
\eea
Here, $r_0$ is the horizon radius and is given by the largest real root of $f(r_0)=0$. Also, note that $M=1+Q^2$. The temperature of the black hole takes the form
\bea\label{temp}
T=\frac{r_0}{4\pi L^2}(3-Q^2)=\frac{\mu}{4\pi}\frac{3-Q^2}{Q}, 
\eea
while its entropy, charge and energy densities are given by 
\bea\label{ece}
s=\frac{2\pi}{\kappa_4^2}\left(\frac{r_0}{L}\right)^2,\qquad \qquad \rho=\frac{2}{\kappa_4^2}\left(\frac{r_0}{L}\right)^2Q,\qquad\qquad \epsilon=\frac{r_0^3}{\kappa_4^2L^4}M,
\eea
respectively \cite{Chamblin:1999tk}. The background is extremal when $Q^2=3$, for which the temperature vanishes but the entropy density remains finite. The background is invariant under flipping the sign of $A_t$. We choose $\mu$ to be positive resulting in $Q$ to be positive. With this convention, $Q=\sqrt{3}$ at extremality. 

The background is dual to a $(2+1)$-dimensional strongly-coupled field theory at finite temperature $T$ and finite charge density $\rho$. The entropy and energy densities of the dual theory are given by $s$ and $\epsilon$ in (\ref{ece}), respectively. Also, the asymptotic value of the bulk gauge field $A_t(\infty)=\mu$ is interpreted in the dual theory as the chemical potential for the (electric) charge density.  Little is known about the details of the dual theory from field theory perspectives. On the other hand, using holography a lot has been learned (especially thermodynamical properties) regarding its strong-coupling behavior; see \cite{Chamblin:1999tk} and its citations.

In this paper we work in the extremal limit, where the dual theory is at zero temperature but finite charge density. We will refer to this dual theory as the boundary field theory. Although the black hole temperature vanishes at extremality, its horizon area remains finite whose dual interpretation is that  the boundary theory has a finite entropy density at zero temperature. 

In the extremal limit, $f(r)$ in the background metric (\ref{metric}) takes the form 
\bea\label{extf}
f(r)=1-4\left(\frac{r_0}{r}\right)^3+3\left(\frac{r_0}{r}\right)^4,
\eea
which has a double zero at the horizon, and can be approximated near that region (to the leading order in $r-r_0$) by 
\bea
f(r)\simeq \frac{6}{r_0^2}(r-r_0)^2.
\eea
The near horizon geometry is $\ads_2\times \mathbb{R}^2$. To see the emergence of this geometry, first change the radial coordinate $r$ to $\eta$ defined by 
\beq\label{definition eta}
r-r_0=\frac{L^2}{6\eta}.
\eeq
There is then a scaling limit \cite{Faulkner:2009wj} in which
\begin{align}\label{mgnhrads2}
ds^2=\frac{L^2}{6\eta^2}\Big(-dt^2+d\eta^2\Big)+\frac{r_0^2}{L^2}\Big(dx^2+dy^2\Big),\qquad\qquad A&=\frac{Q}{6\eta}dt. 
\end{align}
The curvature radius of the $\ads_2$ is $L_2=L/\sqrt{6}$. The radial coordinate is interpreted holographically as the renormalization scale of the dual field theory, and the near horizon region corresponds to the IR limit. This implies that  the $\ads_2\times\mathbb{R}^2$ geometry encodes the IR physics ($\omega\to 0$) of the boundary theory.

On general grounds of holography, one expects the gravity on the $\ads_2$ space to be dual to a CFT$_1$.  This led the authors of \cite{Faulkner:2009wj} to suggest that the $(2+1)$-dimensional boundary field theory (which is dual to the extremal charged $\ads_4$ black hole) flows in the IR to a fixed point described by a CFT$_1$. Following \cite{Faulkner:2009wj} we will refer to this CFT$_1$ as the IR CFT. The details of the $\ads_2/{\rm CFT}_1$ correspondence and how exactly the mapping works are poorly understood. In particular, it is not clear whether the CFT$_1$ of the correspondence represents a conformal quantum mechanics or a chiral sector of a (1+1)-dimensional CFT.  Nevertheless, not knowing the details of the IR CFT and just assuming its existence, the authors of \cite{Faulkner:2009wj} were able to show that the low frequency behavior of some observables in the boundary CFT is encoded in the IR CFT.  In what follows, we consider the charge current and the energy-momentum tensor operators of the boundary field theory and elucidate the role of the IR CFT in determining the low frequency behavior of the retarded Green functions of the two operators.

\section{Gauge Field and Metric Fluctuations}\label{sectionthree}

The charge current operator in the boundary field theory is dual to the fluctuations of the bulk gauge field in the transverse direction while the energy-momentum tensor operator is dual to the fluctuations of the background metric. The calculation of the corresponding retarded Green functions starts with solving the linearized Einstein-Maxwell equations for these fluctuations. 

\subsection{Einstein-Maxwell Equations}

To obtain the linearized Einstein-Maxwell equations,  we first define
\bea\label{fluct}
g_{\mu\nu}= \bar g_{\mu\nu}+h_{\mu\nu}, \qquad A_\mu= \bar A_\mu+a_\mu, 
\eea
where $\bar g_{\mu\nu}$ and  $\bar A_\mu$ represent, respectively, the (extremal) metric and the gauge field of the background, and $h_{\mu\nu}$ and $a_\mu$ are the fluctuations. We choose the so-called radial gauge
\bea\label{rgauge}
a_r=0, \qquad h_{r\nu}=0,
\eea
where $\nu=\{t,x,y,r\}$. 
We proceed by Fourier transforming the fluctuations
\begin{align}\label{foux}
h_{\mu\nu}(t,x,r)\sim e^{-i\omega t} e^{ikx} h_{\mu\nu}(r),\qquad
a_\mu(t,x,r)\sim e^{-i\omega t} e^{ik x} a_{\mu}(r),
\end{align}
where, without loss of generality, we used the rotation invariance in the $(x,y)$ plane to set $k_y=0$ and defined $k_x\equiv k$.  The fluctuations split into decoupled groups depending on whether they are even or odd with respect to parity, $y\to -y$. Accordingly, $h_{ty}$, $h_{xy}$, $a_{y}$ have odd parity while $h_{tt}$, $h_{tx}$, $h_{xx}$, $h_{yy}$, $a_{t}$, $a_{x}$ all have even parity.  In this paper, we study the odd parity modes which translates into the shear and (charge) diffusion modes of the boundary theory. The analysis of the even parity modes (sound modes) will be given elsewhere.

It is more convenient to raise the indices in $h_{ty}$ and $h_{xy}$ (by the background metric $\bar g_{\mu\nu}$) and work with ${h^y}_t$, ${h^x}_y$.  It is also convenient
to define the dimensionless quantities 
\bea\label{dimensionless}
u=\frac{r}{r_0},\qquad\qquad \wn=\frac{\omega}{\mu}, \qquad\qquad \qn=\frac{k}{\mu},
\eea
In this notation, the linearized Einstein-Maxwell equations for the odd parity modes ${h^t}_y$, ${h^x}_y$, $a_{y}$ then read
\begin{align}
&f(u)\Big[u^4{h^y}_t''(u)+ 4u^3{h^y}_t'(u)+\frac{12}{\mu} a'_y(u)\Big]-3\qn\left[\wn {h^x}_y(u)+\qn {h^y}_t(u)\right]=0, \label{eqyt}\\ 
& f(u)\Big[u^4 f(u){h^x}_y''(u)+ \left[u^4f'(u)+4u^3f(u)\right]{h^x}_y'(u)\Big]+3\wn\left[\wn {h^x}_y(u)+\qn {h^y}_t(u)\right]=0,\label{eqxy}\\ 
&f(u)\Big[u^4 f(u) a''_y(u)+ u^2\left[u^2f'(u)+2uf(u)\right]a'_y(u)+\mu u^2{h^y}_t'(u)\Big]
+3\left[\wn^2-f(u)\qn^2\right]a_y(u) =0,\label{eqy}
\end{align}
where $f(u)$ is given in (\ref{extf}). There is also a constraint which comes from the $yu$-component of the linearized Einstein equations given by 
\bea\label{cons}
u^4\wn~{h^y}_t'(u)+u^4 f(u)\qn~{h^x}_y'(u)+\frac{12}{\mu} \wn~a_y(u)=0.
\eea
As we will see later in more detail, this constraint, evaluated asymptotically, encodes a Ward identity in the dual boundary theory, as one should expect.

\subsection{Gauge Invariant Modes}

Although working in the radial gauge \eqref{rgauge} is convenient, it does not completely fix the gauge freedom of $a_\mu$ and $h_{\mu\nu}$. In the present case, it is easy to identify the gauge invariant combinations of fields.


For the metric and gauge field fluctuations of our background, there is a U(1) gauge transformation $A_{\mu} \rightarrow {\tilde A}_{\mu}=A_\mu+ \nabla_{\mu}\chi$, according to which $f_{\mu\nu} = \partial_{\mu}a_{\nu} - \partial_{\nu}a_{\mu}$ is invariant. There are also gauge transformations associated with diffeomorphism under an arbitrary vector field $\xi$, implying that the fluctuations $\{\tilde h_{\mu\nu}, {\tilde a}_{\rho}\}$ and $\{h_{\mu\nu}, a_{\rho}\}$ are equivalent if they are related as \cite{Wald:1984rg}
\begin{align}\label{residual gauge}
\tilde h_{\mu\nu} &= h_{\mu\nu} - \left(\xi^{\lambda}\nabla_{\lambda}\bar{g}_{\mu\nu} + \bar{g}_{\lambda \nu}\nabla_{\mu}\xi^{\lambda} + \bar{g}_{\mu\lambda}\nabla_{\nu}\xi^{\lambda}\right)= h_{\mu\nu} - \left(\bar\nabla_{\mu}\xi_{\nu} + \bar\nabla_{\nu}\xi_{\mu}\right),\\
\tilde a_{\mu} &= a_{\mu}-\left(\xi^{\nu}\nabla_{\nu}\bar{A}_{\mu} +\bar{A}_{\nu}\nabla_{\mu}\xi^{\nu}\right) = a_{\mu}-\left(\xi^{\nu}\partial_{\nu}\bar{A}_{\mu} +\bar{A}_{\nu} \partial_{\mu}\xi^{\nu}\right),
\end{align}
where $\nabla$ is any torsionless connection and $\bar \nabla$ is the connection associated with $\bar g_{\mu \nu}$. It is important to notice that these gauge transformations are not changes of the coordinates. In particular, they affect only the fields $h_{\mu\nu}$ and $a_{\mu}$, while all the other tensors, vectors, etc. remain invariant. 

In this paper, the fluctuations of interest are ${h^y}_t$, ${h^x}_y$ and $a_y$. 
While $a_{y}$ is invariant under the residual gauge transformations, ${h^y}_t$ and ${h^x}_y$ transform according to ${h^y}_t \rightarrow {h^y}_t + i\mu\wn \xi^{y}$ and ${h^x}_y\rightarrow {h^x}_y- i\mu\qn\xi^{y}$, respectively. Thus
\begin{align}\label{ginvone}
X(u)&=\qn {h^y}_t(u) + \wn {h^x}_y(u),\\
Y(u)&= a_{y}(u), \label{ginvtwo}
\end{align}
are invariant combinations.

Note that at $\qn=0$ the symmetries are enhanced, and ${h^x}_y(u)$ is gauge invariant. For this special case, as can be seen in equation \eqref{eqxy},  ${h^x}_y(u)$ decouples from the rest, while equations \eqref{eqy} and \eqref{cons} together give a decoupled equation for $a_y(u)$. The decoupled equations read
\begin{align}
&u^4 f(u){h^x}_y''(u)+ \left[u^4f'(u)+4u^3f(u)\right]{h^x}_y'(u)+\frac{3\wn^2}{f(u)} {h^x}_y(u)=0,\label{hxyzeroq}\\
&u^2 f(u)a''_y(u)+ \left[u^2f'(u)+2uf(u)\right]a'_y(u)+\frac{1}{u^2}\left(\frac{3\wn^2}{f(u)}-\frac{12}{u^2}\right)=0\label{ayzeroq}.
\end{align}
Equations \eqref{hxyzeroq} and \eqref{hxyzeroq} were studied in \cite{Edalati:2009bi} and the retarded Green functions  of the dual operators, $T_{xy}$ and $J_y$, at $\qn=0$ and small $\wn$ were calculated. 
For the most part in this paper, we consider $\qn$ to be finite and non-zero. 

Even at finite $\qn$ and $\wn$, the Einstein-Maxwell shear channel equations can be decoupled in terms of a pair of  ``master fields" \cite{Kodama:2003kk}. This is remarkable, since in the boundary field theory it corresponds to diagonalizing the entire renormalization group flow.
The master fields are
\begin{align}\label{phipmdef}
\Phi_{\pm}(u)&=-\mu \frac{\qn f(u)u^3}{\wn^2-f(u)\qn^2}X'(u)-\frac{6}{u}\left[\frac{2f(u)\qn^2}{\wn^2-f(u)\qn^2}+u\left(1\pm\sqrt{1+\qn^2}\right)\right]Y(u).
\end{align}
and they satisfy the decoupled equations 
\begin{align}\label{phipmeq}
\left[u^2f(u)\Phi^{\prime}_{\pm}(u)\right]^{\prime}+\left[uf'(u)+\frac{3}{u^2f(u)}\left(\wn^2-f(u)\qn^2\right)-\frac{6}{u^3}\left(1\pm\sqrt{1+\qn^2}\right)\right]\Phi_{\pm}(u)&=0.
\end{align}
As we will discuss in the next section, it is most convenient to organize numerical calculations in terms of these modes.

\section{Retarded Green Functions and Criticality}\label{sectionfour}

In this section, we first obtain formal expressions for the retarded Green functions of the vector current and energy momentum tensor operators of the boundary theory. By formal expressions we mean that we write formulas relating the above-mentioned retarded Green functions to $\hat\Pi_{\pm}$, to be introduced below.

Consider first the solution of the equations \eqref{eqyt}, \eqref{eqxy} and \eqref{eqy} asymptotically, for $u\to\infty$. For each of the fields in this problem, there is a constant asymptotic value that we denote by a hat.
The solutions take the form
\begin{align}
{h^x}_y(u\to\infty)&= {{\hat h}^x}_{~y}+\frac32 \frac{\wn}{u^2}\left(\wn{{\hat h}^x}_{~y}+\qn {{\hat h}^y}_{~t}\right)+\frac{{{\pi}^x}_{y}}{u^3}+\ldots,\\
{h^y}_t(u\to\infty)&= {{\hat h}^y}_{~t}-\frac32 \frac{\qn}{u^2}\left(\wn{{\hat h}^x}_{~y}+\qn {{\hat h}^y}_{~t}\right)+\frac{{{\pi}^y}_{t}}{u^3}+\ldots,\\
a_y(u\to\infty)&= {\hat a}_y+\frac{\pi_y}{u}-\frac32 \frac{(\qn^2-\wn^2)}{u^2}\left(\wn{{\hat h}^x}_{~y}+\qn {{\hat h}^y}_{~t}\right)+\ldots.
\end{align}
In these expressions, all quantities are understood to be functions of $\wn$ and $\qn$ even if not explicitly indicated.
Here, ${{\hat h}^x}_{~y},  {{\hat h}^y}_{~t}$ and ${\hat a}_y$ are regarded, as usual, as the sources for the corresponding operators in the dual field theory, while ${{\pi}^x}_{y},  {{\pi}^y}_{t}$ and ${\pi}_y$ are the corresponding one-point functions of those operators. The latter are understood to be functions of the sources, and that functional dependence, given the on-shell action, encodes the correlation functions of the dual field theory.

The renormalized action in our case is given by
\begin{align}\label{action}
2\kappa_{4}^{2}\,S_{\rm ren} &=\int d^{4}x \sqrt{-g}\left(R+\frac{6}{L^2}- L^{2}F_{\mu\nu}F^{\mu\nu}\right)\nonumber\\
&-\int_{\partial M}d^{3}x\,\sqrt{|\gamma|}2K
- \frac{4}{L}\int_{\partial M}d^{3}x\,\sqrt{|\gamma|} - L\int_{\partial M}d^{3}x\,\sqrt{|\gamma|}\ {}^{(3)}R,
\end{align}
where $\gamma_{\mu\nu}$,  $K$ and ${}^{(3)}R$ are, respectively, the induced metric, the trace of the second fundamental form and the intrinsic curvature of the 3-dimensional boundary $\partial M$.
The counterterm proportional to ${}^{(3)}R$ removes a linear divergence in the on-shell action for the Reissner-Nordstr\"{o}m AdS$_4$  background, with no contribution to its finite part. In order to extract Green functions that we are interested in holographically, it is enough to  consider the (gravity) action up to quadratic terms in the fluctuations. One finds 
\begin{align}\label{onshellaction}
S_{\rm ren}=\lim_{u\to\infty}C\int d^3x~&\Big[u^4 {h^y}_t {h^y}_t'-u^4f(u){h^x}_y {h^x}_y'+4u^3\left(1-f(u)^{-1/2}\right)\left({h^y}_t {h^y}_t-f(u){h^x}_y {h^x}_y\right)\nonumber\\
&-u^4f'(u){h^x}_y {h^x}_y-\frac{12}{\mu^{2}}\left(\mu a_y {h^y}_t+u^2f(u)a_ya_y'\right)\nonumber\\
&-\frac{3}{\mu^{2}}uf(u)^{-1/2}\left(\partial_{t}{h^{x}}_{y}-\partial_{x} {h^{y}}_{t}\right)^2\Big],
\end{align}
where $C=r_0\mu^2/(12\kappa_4^2)$. 

The holographic prescription for evaluating the matrix of retarded Green functions when the operators mix was given in \cite{Son:2006em} following \cite{Son:2002sd, Herzog:2002pc}\footnote{See also \cite{Amado:2009ts, Kaminski:2009dh} for a recent more systematic treatment of holographically computing Green functions in cases where the boundary theory operators mix.}. Because we have rotational invariance in the spatial plane, we will always take the $y$-component of momentum to vanish, and hence it is most convenient to Fourier transform the $x$-direction only. Having imposed infalling boundary condition at the horizon, one writes the renormalized on-shell action in the form  
\begin{align}
S_{\rm o.s.}=\int dy \int \frac{d\wn d\qn}{(2\pi)^2} ~\hat \phi_i (\wn,\qn) {\cal F}_{ij}(\wn,\qn)\hat \phi_j (-\wn,-\qn),
\end{align}
where $\hat\phi_i=\{{\hat h^y}_{~t}, {\hat h^x}_{~y}, \hat a_y\}$ in our case. The prescription of \cite{Son:2006em}  then reads 
\begin{align}\label{Greendef}
G_R(\wn,\qn)=\left\{\begin{array}{cl}
-2{\cal F}_{ij}(\wn,\qn) & i=j \\
-{\cal F}_{ij}(\wn,\qn)& i\neq j. 
\end{array}\right.
\end{align}

In momentum space, \eqref{onshellaction} evaluates to
\begin{align}
S_{ren}=3\mu^2C\int dy\int \frac{d\wn d\qn}{(2\pi)^2}~&\Big[ {\pi^x}_y(\wn,\qn){\hat h^x}_{~y}(-\wn,-\qn)-{\pi^y}_t(\wn,\qn){\hat h^y}_{~t}(-\wn,-\qn)\nonumber\\
&+\frac{4}{\mu^2}\pi_y(\wn,\qn)\hat a_y(-\wn,-\qn)+\ldots
\Big],
\end{align}
where the ellipsis contains terms which contribute at most to contact terms in two-point functions.

Now, as we recalled above, the $\{  {\pi^x}_y,{\pi^y}_t,\pi_y\}$ are regarded as functions of $\{{\hat h^y}_{~t}, {\hat h^x}_{~y}, \hat a_y\}$, and for present purposes, they can be regarded as linear functions. Also, by assumption, none of the corresponding operators have one-point functions when the sources are removed.  The vevs are not all independent in this case though. Indeed, the constraint (\ref{cons}) gives one more piece of information
\begin{equation}\label{wardonepoint}
\qn {\pi^x}_{y}+\wn {\pi^y}_{t}=\frac{4}{\mu} \wn \hat a_y
\end{equation}
This equation is easily understood in the boundary theory. The diffeomorphism Ward identity \cite{Herzog:2009xv} takes the form
\begin{equation}
-g_{\nu\lambda}(x) \nabla_\mu\langle \hat T^{\mu\nu}(x)\rangle+ F_{\lambda\mu}(x)\langle \hat J^\mu(x)\rangle=0.\label{diffward}
\end{equation}
whose $y$-component evaluates to (\ref{wardonepoint}) at linearized order given the source $\hat a_y$,  and our notation.

Thus, there are really only two independent vevs here, in accordance with the fact that the bulk problem can be reduced to the two gauge-invariant fields. What is more, the system is in fact diagonalized by passing to the master fields. Asymptotically, we write the master fields as
\begin{align}\label{asymPhipm0}
\Phi_{\pm}(u\to \infty)\simeq  \hat\Phi_{\pm}\left(1+\frac{\hat\Pi_{\pm}}{u}+\ldots\right),
\end{align}
where  $\hat\Phi_{\pm}$ and $\hat\Pi_{\pm}$ are functions of $\wn$ and $\qn$, and we have in mind  that the master fields satsify infalling boundary conditions at the horizon.

By comparing asymptotic expansions, one easily deduces that 
\begin{align}
\hat\Phi_{\pm}=3\mu\qn\left( \wn{{\hat h}^x}_{~y} +\qn{{\hat h}^y}_{~t} \right)-6\left(1\pm\sqrt{1+\qn^2}\right)\hat a_y,
\end{align}
while
\begin{equation}
\hat\Phi_{\pm}\hat\Pi_{\pm} = 3\frac{\qn}{\wn}{\pi^x}_y-6\left( 1\pm \sqrt{1+\qn^2}\right)\pi_y,
\end{equation}
where we have used (\ref{wardonepoint}) to simplify. It should be clear then that the on-shell action can be written as a function of the sources  $\{{\hat h^y}_{~t}, {\hat h^x}_{~y}, \hat a_y\}$ and $\hat\Pi_{\pm}$.
The latter two functions (of $\wn$ and $\qn$) must be determined numerically. The matrix of retarded two-point functions in the shear channel are then determined by these two functions. This is in fact consistent with the requirements of the Ward identities.

Returning to the on-shell action, using the holographic prescription given above, we deduce the retarded correlators of $\hat J_y$, $\hat T_{xy}$ and $\hat T_{yt}$
\begin{align}
G_{yt,yt}(\wn,\qn)=&\mu^2\qn^2G_1(\wn,\qn),\label{Gytyt}\\
G_{yt,xy}(\wn,\qn)=&-\mu^2\qn\wn G_1(\wn,\qn),\\
G_{xy,xy}(\wn,\qn)=&\mu^2\wn^2G_1(\wn,\qn),\\
G_{yt,y}(\wn,\qn)=&-\frac12\mu\qn^2 \left[G_1(\wn,\qn)+G_2(\wn,\qn)\right],\\
G_{xy,y}(\wn,\qn)=&\frac12\mu\qn\wn\left[G_1(\wn,\qn)+G_2(\wn,\qn)\right],\\
G_{y,y}(\wn,\qn)=&-4G_2(\wn,\qn),\label{Gyy}
\end{align}
where
\begin{align}\label{G1}
G_1(\wn,\qn)&\equiv\frac{3C}{\sqrt{1+\qn^2}}\left[\left(1-\sqrt{1+\qn^2}\right)\hat\Pi_+-\left(1+\sqrt{1+\qn^2}\right)\hat\Pi_-\right],\\
G_2(\wn,\qn)&\equiv\frac{3C}{\sqrt{1+\qn^2}}\left[\left(1+\sqrt{1+\qn^2}\right)\hat\Pi_+-\left(1-\sqrt{1+\qn^2}\right)\hat\Pi_-\right].\label{G2}
\end{align}
We have implicitly assumed here that the bulk equations of motion are to be solved by placing infalling boundary conditions on $\Phi_\pm(u)$.
It can easily be verified that the prescription  \eqref{Greendef} implies that $G_{xy,yt}(\wn,\qn)=G_{yt,xy}(\wn,\qn)$, $G_{y,xy}(\wn,\qn)=G_{xy,y}(\wn,\qn)$, and up to a contact term $G_{y,yt}(\wn,\qn)=G_{yt,y}(\wn,\qn)$.

\subsection{Frequency Expansions of $\hat\Phi_{\pm}$ and $\hat\Pi_{\pm}$}

The expressions we derived above for the retarded Green functions are not very informative unless   $\hat\Pi_{\pm}(\wn,\qn)$ is known. One thing that can be done semi-analytically is to develop a series expansion in $\wn$. This is an extension of our earlier results to finite momentum. This analysis makes use of the fact that the near horizon geometry is ${\rm AdS}_2\times\mathds{R}^2$.  

 Let us expand  $\Phi_{\pm}(u)$ in a power series in $\wn$:
\begin{align}\label{outerPhi}
\Phi_{O\pm}(u)&=\Phi^{(0)}_{O\pm}(u)+\wn \Phi^{(1)}_{O\pm}(u)+\wn^2 \Phi^{(2)}_{O\pm}(u)+\ldots,
\end{align}
where the subscript denotes the outer region (see the discussion below). Since the black hole background is extremal, taking the limit of $\wn \to 0$ becomes somewhat more subtle near the horizon. This is due to the fact that in the extremal case $f(u)$ has a double zero at the horizon. A way to handle this issue was explained in \cite{Faulkner:2009wj}. Basically, one realizes that near the horizon the equations for $\Phi_{\pm}$ in \eqref{phipmeq} organize themselves as functions of $\zeta\equiv \omega\eta$, where, using \eqref{definition eta} and \eqref{dimensionless}, we have 
\beq\label{inner}
u=1+\frac{\wn}{\sqrt{12}\zeta}.
\eeq
Since the coordinate $\zeta$ is the suitable radial coordinate for the $\ads_2$ part of the near horizon geometry, the $\wn$ expansion of  $\Phi_{\pm}$ in that region can be written 
\begin{align}\label{innerPhi}
\Phi_{I\pm}(\zeta)&=\Phi^{(0)}_{I\pm}(\zeta)+\wn \Phi^{(1)}_{I\pm}(\zeta)+\wn^2 \Phi^{(2)}_{I\pm}(\zeta)+\ldots, 
\end{align}
where the subscript  $I$ denotes the inner region. 
Having implemented the (infalling) boundary condition at the horizon  \cite{Son:2002sd,Herzog:2002pc} in the $\zeta$ coordinate, one then matches the inner and outer expansions in the ``matching region" where the  $\zeta\to 0$ and $\wn/\zeta\to 0$ limits are taken.  Since the differential equations (\ref{phipmeq}) are linear,  and we also require that the solutions for the higher order terms in the expansions (\ref{outerPhi}) and (\ref{innerPhi}) do not include terms proportional to  the zeroth-order solutions near the matching region, we just need to match $\Phi^{(0)}_{I\pm}(\zeta)$  to $\Phi^{(0)}_{O\pm}(u)$ near that region \cite{Faulkner:2009wj} at leading order. 


\subsubsection{Inner Region}

Substituting \eqref{innerPhi} in \eqref{phipmeq}, we find that the leading terms satisfy
\begin{align}\label{leadinglinner}
-\Phi^{(0)\prime\prime}_{I\pm}(\zeta)+\left(\frac{\qn^2+2\pm2\sqrt{1+\qn^2}}{2\zeta}-1\right)\Phi^{(0)}_{I\pm}(\zeta)=0. 
\end{align}
Equations \eqref{leadinglinner} are identical to the equations of motion for massive scalar fields with effective AdS$_2$ masses 
\begin{align}
m_\pm ^2 L_2^2&=1+\frac{\qn^2}{2}\pm\sqrt{1+\qn^2}.
\end{align}
Thus $\Phi^{(0)}_{I\pm}(\zeta)$ source operators ${\cal O}_{\pm}$ in the IR CFT with conformal dimensions $\delta_{\pm}=\nu_{\pm}+\frac{1}{2}$ where 
\begin{align}\label{nupm}
\nu_{\pm}=\frac{1}{2}\sqrt{1+4m_\pm ^2 L_2^2}=\frac{1}{2}\sqrt{5+2\qn^2\pm 4\sqrt{1+\qn^2}}.
\end{align}
Having imposed the infalling boundary conditions at $\zeta\to\infty$, $\Phi^{(0)}_{I\pm}(\zeta)$ take the following form  near the matching region (where one takes the $\zeta\to 0$ and $\wn/\zeta\to 0$ limits)
\begin{align}\label{inzerothphipm}
\Phi^{(0)}_{I\pm}(u\to1)=\left[(u-1)^{-\frac{1}{2}+\nu_{\pm}}+{\cal G}_{\pm}(\wn)(u-1)^{-\frac{1}{2}-\nu_{\pm}}\right].
\end{align}
Note that in writing \eqref{inzerothphipm} we used  \eqref{inner} to express $\zeta$ in terms of $u$. Also, we chose a specific normalization for $\Phi^{(0)}_{I\pm}$. Such a choice does not affect the calculation of the boundary theory Green functions. In \eqref{inzerothphipm}, ${\cal G}_{\pm}(\wn)$ denote the retarded Green functions of the aforementioned  IR CFT operators ${\cal O}_{\pm}$, and are given by \cite{Faulkner:2009wj, Edalati:2009bi}
\begin{align}\label{tworetgreen}
{\cal G}_\pm(\wn)=-2\nu_\pm e^{-i\pi\nu_\pm}\frac{\G(1-\nu_\pm)}{\G(1+\nu_\pm)}\left(\frac{\wn}{2}\right)^{2\nu_\pm}.
\end{align}
The analytic structure of $\cal G_\pm(\wn)$ is the same as the retarded Green function of an IR CFT scalar operator discussed in \cite{Faulkner:2009wj}. Namely, for generic values of momentum $\qn$, there is a branch point at $\wn=0$, and moreover the branch cut is chosen to extend along the negative imaginary axis in the  complex $\wn$-plane. 

\subsubsection{Outer Region and Matching}
In the outer region, the equations for $\Phi^{(0)}_{O\pm}(u)$ are obtained by setting $\wn=0$ in \eqref{phipmeq}. It is easy to show that near the matching region,  the solutions for $\Phi^{(0)}_{O\pm}(u)$ are given by a linear combination of $(u-1)^{-\frac{1}{2}+\nu_{\pm}}$ and $(u-1)^{-\frac{1}{2}-\nu_{\pm}}$. For ease of notation, we define
\begin{align}
\eta^{(0)}_{\pm}(u)=(u-1)^{-\frac{1}{2}+\nu_{\pm}}+\ldots, \qquad \qquad \xi^{(0)}_{\pm}(u)=(u-1)^{-\frac{1}{2}-\nu_{\pm}}+\ldots, \qquad u\to 1.
\end{align}
Then, matching  $\Phi^{(0)}_{O\pm}(u)$ to \eqref{inzerothphipm},  we obtain
\begin{align}
\Phi^{(0)}_{O\pm}(u)=\left[\eta^{(0)}_{\pm}(u)+{\cal G}_\pm(\wn)\xi^{(0)}_{\pm}(u)\right].
\end{align}

The discussion here follows that of \cite{Faulkner:2009wj}. To higher orders, we can write
\begin{align}
\eta_{\pm}(u)&=\eta^{(0)}_{\pm}(u)+\wn \eta^{(1)}_{\pm}(u)+\wn^2 \eta^{(2)}_{\pm}(u)+\ldots,\\
\xi_{\pm}(u)&=\xi^{(0)}_{\pm}(u)+\wn \xi^{(1)}_{\pm}(u)+\wn^2 \xi^{(2)}_{\pm}(u)+\ldots, 
\end{align}
where $\eta^{(n>0)}_{\pm}(u)$ and $\xi^{(n>0)}_{\pm}(u)$ are obtained demanding that in the $u\to 1$ limit, they are distinct from $\eta^{(0)}_{\pm}(u)$ and $\xi^{(0)}_{\pm}(u)$, respectively. Thus, 
\begin{align}
\Phi_{O\pm}(u)=\left[\eta_{\pm}(u)+{\cal G}_\pm(\wn)\xi_{\pm}(u)\right].
\end{align}
Near $u\to \infty$, one can expand $\eta^{(n)}_{\pm}(u)$ and $\xi^{(n)}_{\pm}(u)$ as follows (here $n\geq 0$)
\begin{align}
\eta^{(n)}_{\pm}(u\to\infty)&=a^{(n)}_{\pm} \Big(1+\ldots\Big)+b^{(n)}_{\pm} \frac{1}{u}\Big(1+\ldots\Big),\\
\xi^{(n)}_{\pm}(u\to\infty)&=c^{(n)}_{\pm} \Big(1+\ldots\Big)+d^{(n)}_{\pm}\frac{1}{u}\Big(1+\ldots\Big),
\end{align}
where the coefficients $a^{(n)}_\pm, b^{(n)}_\pm, c^{(n)}_\pm$ and  $d^{(n)}_\pm$  are all functions of $\qn$. So, asymptotically, we obtain
\begin{align}\label{xpm}
{\hat\Phi}_{\pm} &=\left[a^{(0)}_\pm+\wn a^{(1)}_\pm+{\cal O}\left(\wn^2\right)\right]+ {\cal G}_\pm(\wn)\left[ c^{(0)}_\pm+\wn c^{(1)}_\pm+{\cal O}\left(\wn^2\right)\right],\\
{\hat\Phi}_{\pm}{\hat\Pi}_{\pm}&=\left[b^{(0)}_\pm+\wn b^{(1)}_\pm+{\cal O}\left(\wn^2\right)\right]+ {\cal G}_\pm(\wn)\left[ d^{(0)}_\pm+\wn d^{(1)}_\pm+{\cal O}\left(\wn^2\right)\right].\label{ypm}
\end{align}
By virtue of  \eqref{Gytyt}--\eqref{Gyy}, these determine the frequency dependence of the retarded Green functions in the boundary theory, for small $\wn$.

\subsection{Criticality: Emergent IR Scaling}

Although the Green functions \eqref{Gytyt}--\eqref{Gyy} can be numerically computed  by computing the coefficients $a^{(n)}_\pm, b^{(n)}_\pm, c^{(n)}_\pm, d^{(n)}_\pm$, it turns out  that  one can analyze the low frequency behavior of the Green functions without knowing these coefficients explicitly. For the cases of scalar and spinor operators, such analyses  were performed in \cite{Faulkner:2009wj} where a variety of emergent IR phenomena were observed. In this subsection, we perform similar analyses of  the retarded Green functions of the vector current  and the energy-momentum tensor operators of the boundary theory at low frequency and determine their emergent IR behaviors. As we will see momentarily, one of the reasons which makes the low frequency analyses of the above-mentioned retarded Green functions non-trivial and interesting is the fact that there are now two IR CFT Green functions ${\cal G}_{\pm}(\wn)$ both of which can potentially play a role. Not surprisingly, the mixing of ${\cal G}_{\pm}(\wn)$ in the expressions \eqref{Gytyt}--\eqref{Gyy} for the boundary theory Green functions is due to the coupled nature of the electromagnetic and gravitational perturbations we are considering in the charged (extremal) black hole background.

As defined in \eqref{nupm}, $\nu_\pm$ are real implying that $a^{(n)}_\pm, b^{(n)}_\pm, c^{(n)}_\pm$ and $d^{(n)}_\pm$ are all real. Thus the complex part of $\hat\Pi_{\pm}$ comes entirely from the IR CFT Green functions ${\cal G}_\pm (\wn)$. Suppose for generic momentum $\qn$, the product $a^{(0)}_{+}a^{(0)}_{-}$ is non-vanishing. Expanding  \eqref{G1} and \eqref{G2} for small $\wn$, and noticing that  $2\nu_{+}>3$ while $2\nu_{-}>1$, one then deduces that ${\rm Im} ~G_1(\wn,\qn)$ and ${\rm Im} ~G_2(\wn,\qn)$ exhibit scaling behavior at small frequency:
\begin{align}\label{sfImG1}
{\rm Im} ~G_1(\wn,\qn)&\simeq\frac{3C}{\sqrt{1+\qn^2}}\left(1+\sqrt{1+\qn^2}\right)e_{0}(\qn)~{\rm Im}~ {\cal G}_{-}(\wn)\left[1+\ldots\right]\propto \wn^{2\nu_{-}},\\
{\rm Im} ~G_2(\wn,\qn)&\simeq\frac{3C}{\sqrt{1+\qn^2}}\left(1-\sqrt{1+\qn^2}\right)e_{0}(\qn)~{\rm Im}~ {\cal G}_{-}(\wn)\left[1+\ldots\right]\propto \wn^{2\nu_{-}},\label{sfImG2}
\end{align}
where
\begin{align}
e_{0}(\qn)=\frac{b^{(0)}_{-}}{a^{(0)}_{-}}\left(\frac{c^{(0)}_{-}}{a^{(0)}_{-}}-\frac{d^{(0)}_{-}}{b^{(0)}_{-}}\right).
\end{align}
Substituting \eqref{sfImG1} and \eqref{sfImG2} into \eqref{Gytyt}--\eqref{Gyy}, the spectral functions (the imaginary part of the aforementioned retarded Green functions) of the vector current $\hat J_y$ and the energy-momentum tensor operators $\hat T_{yt}$ and $\hat T_{xy}$ of the boundary field theory show scaling behavior at small frequency
\begin{equation}
\begin{array}{lll}
{\rm Im}~G_{ytyt}(\wn,\qn)\propto \wn^{2\nu_{-}},&\qquad  {\rm Im} ~G_{xyxy}(\wn,\qn)\propto \wn^{2+2\nu_{-}},&\qquad {\rm Im} ~G_{ytxy}(\wn,\qn)\propto \wn^{1+2\nu_{-}},\\
{\rm Im} ~G_{yty}(\wn,\qn)\propto \wn^{2\nu_{-}},&\qquad {\rm Im} ~G_{xyy}(\wn,\qn)\propto \wn^{1+2\nu_{-}},&\qquad {\rm Im} ~G_{yy}(\wn,\qn)\propto \wn^{2\nu_{-}}.
\end{array} 
\end{equation}

Note that the low-frequency scaling behavior of the spectral functions above is emergent meaning that  it is a consequence of the fact that the near horizon region of the background geometry (which translates into the IR physics of the boundary theory) contains an AdS$_2$ factor. Although $e_0$ depends on details of the background geometry, the scaling behavior $\wn^{2\nu_{-}}$ does not change  by changing the geometry in the outer region as long as the AdS$_2$ part of the near horizon geometry is kept intact.  Also, note that because $\nu_\pm$ are strictly real, the retarded Green functions \eqref{Gytyt}--\eqref{Gyy} do not exhibit the log-periodicity behavior observed in \cite{Liu:2009dm, Faulkner:2009wj} for the cases of charged scalar and spinor operators. 

Of course, it is also of interest to consider the physical poles and branch points of the correlation functions at arbitrary $\qn$ and $\wn$. We turn to a study of that problem in the following sections.

\section{Retarded Green Functions and Quasinormal Modes}\label{sectionfive}

The expressions \eqref{Gytyt}--\eqref{Gyy} for the retarded Green functions of the boundary field theory indicate that they generically have poles  whenever $\hat\Phi_{\pm}(\wn,\qn)=0$.  We denote by $\wn^{\pm}_n(\qn)$ the frequencies at which  $\hat\Phi_{\pm}(\wn,\qn)$ vanish. 
In order to obtain our boundary theory Green functions holographically, we imposed infalling boundary condition on $\Phi_{\pm}$ at the horizon.  By further imposing $\Phi_{\pm}$ to vanish asymptotically we are essentially  solving for their quasinormal modes (QNMs) in the extremal Reissner-Nordstr\"{o}m AdS$_{4}$ black hole  background. 

In the context of the AdS/CFT correspondence, the connection between the QNMs of gravitational backgrounds and the singularities of the retarded Green functions of  the dual boundary theories was first suggested in \cite{Birmingham:2001pj}. It was then argued in  \cite{Son:2002sd}  that the quasinormal frequencies of the decoupled fluctuations of a black hole background coincide with the poles in the retarded Green functions of the corresponding dual operators in the boundary theory. The authors of  \cite{Kovtun:2005ev} then argued that in cases where the fluctuations are coupled, in order for the quasinormal frequencies to coincide with the poles in the retarded Green functions of the dual operators, one should consider the gauge invariant combinations as the right set of variables and impose on them infalling boundary condition at the horizon together with an asymptotic normalizable boundary condition. In our case, as it is seen from the linearized Einstein-Maxwell equations \eqref{eqyt}--\eqref{cons}, the electromagnetic and gravitational perturbations of the background are certainly coupled. In section \ref{sectionthree} we constructed the gauge invariant combinations $X(u)$ and $Y(u)$ in terms of which  the linearized Einstein-Maxwell equations are still coupled. We decoupled the equations for the gauge-invariant modes by introducing new variables $\Phi_{\pm}(u)$. These variables  are themselves gauge invariant by virtue of their definition \eqref{phipmdef} in terms of $X(u)$ and $Y(u)$. Thus, with $\Phi_{\pm}(u)$ being gauge invariant and satisfying decoupled second-order differential equations, we can analyze their QNMs. 


To our understanding of  the literature, the electromagnetic and gravitational QNMs  of the extremal Reissner-Nordstr\"{o}m AdS$_{4}$ black hole have not been studied before\footnote{The QNMs of  the four-dimensional asymptotically flat extremal Reissner-Nordstr\"{o}m black hole were studied in \cite{Onozawa:1995vu}. Ref. \cite{Berti:2003ud} analyzed the scalar, electromagnetic and gravitational QNMs of non-extremal Reissner-Nordstr\"{o}m AdS$_{4}$ black hole. Recently, the authors of \cite{Denef:2009yy} studied the QNMs of a massive, charged scalar field in the background of the electrically, as well as the dyonic extremal Reissner-Nordstr\"{o}m AdS$_{4}$ black hole.}, which makes finding these QNMs an interesting problem in its own. Ref. \cite{Berti:2009kk} is a nice review of QNMs from both general relativity and the AdS/CFT perspectives. 

There are only very few backgrounds whose QNMs  are known analytically. Although there are semi-analytic methods for calculating small or large (magnitudes of) quasinormal frequencies, one is usually forced to do numerics in order to extract the generic values of such frequencies. We refer the reader to \cite{Berti:2009kk} for an extensive list of analytic and numerical methods available in the literature on how to compute the QNMs of different backgrounds. In this section, 
we numerically compute the quasinormal frequencies of the  (shear channel) electromagnetic and gravitational perturbations of the extremal Reissner-Nordstr\"{o}m AdS$_{4}$ black hole. Along the way, we find it useful to compare our zero-temperature numerical results with the analogous QNMs of the non-extremal Reissner-Nordstr\"{o}m AdS$_{4}$ black hole background.  Although, the results in the non-extremal case are not new, we present them here for the purpose of comparison and to study the $T\to 0$ limit (at fixed $\mu$). By doing so, we will see more clearly how the analytic structure at $T=0$ is related to that of finite temperature.

\subsection {Generalities} 

Before numerically computing the quasinormal frequencies  for our problem, we discuss some general features that one expects to observe.
\subsubsection{Branch Cut}

For small $\wn$, the analytic structure is determined by the small frequency analysis that we presented
earlier, pertaining to the near-horizon geometry. 
For generic values of $\qn$,  ${\cal G}_{\pm}(\wn)$  are multi-valued and  the boundary theory Green functions are then generically multi-valued as well. When $\qn=0$, $2\nu_{\pm}$ are integers: indeed $2\nu_{+}=3$ and $2\nu_{-}=1$. In this case, the multi-valuedness of the Green functions is due to the existence of terms like $\wn^{m}\log{\wn}$ in their expressions, where $m$ is a positive integer, in the extremal case.\footnote{As we will recall below, such logs are absent at finite temperature, and the cut resolves into a series of Matsubara poles along the negative imaginary axis.} For example, one can show that  at small $\wn$
\begin{align}
G_{xy,xy}(\wn,\qn=0)\propto \frac{i \wn}{1+c_1 \wn\log{\wn}+c_2 i\wn},
\end{align}
where $c_1$ and $c_2$ are numerical factors. Thus, the QNMs of $\Phi_{\pm}$ will also be multi-valued (in the extremal case). There is another, perhaps, more direct argument for the existence of the branch cut in the QNMs of $\Phi_{\pm}$, which is based on analyzing the ``tail" of the $\Phi_{\pm}$ fluctuations in the extremal black hole background \cite{Ching:1995tj}. An argument of this type was used in \cite{Denef:2009yy} to show the existence of the branch cut in the QNMs of a chargeless massless scalar in the extremal Reissner-Nordstr\"{o}m AdS$_{4}$ black hole background, and  can easily be adopted to our case, as well. To that end, define the (dimensionless) tortoise coordinate $u_{*}$ via
\begin{align}\label{tortoise}
\frac{du_*}{du}=\frac{1}{u^2 f(u)}.
\end{align}
The asymptotic boundary of the background in the tortoise coordinate is obtained by taking  $u_*\to 0$ whereas $u_*\to-\infty$ is  the near horizon region. Using the tortoise coordinate, we can write the equations in \eqref{phipmeq}  in the form of Schr\"{o}dinger equations (where we think of $\qn$ as being fixed)
\begin{align}\label{Scheq}
\Phi^{\prime\prime}_{\pm}(u_*)+ \left[3\wn^2 -V_{\pm}(u_*)\right]\Phi_{\pm}(u_*)=0, 
\end{align}
where the potentials $V_{\pm}(u_*)$ are given by 
\begin{align}
V_{\pm}(u_*)= \frac{6}{u}f(u)\left(1\pm\sqrt{1+\qn^2}\right)+3\qn^2f(u)-u^3 f(u)f'(u).
\end{align}
It is understood that in the above expressions for $V_{\pm}(u_*)$, one should use  \eqref{tortoise}  to express $u$ in terms of $u_*$. Near the horizon, one easily finds
\begin{align}
V_{\pm}(u_*\to-\infty)\sim\frac{1}{ u_*^3}.
\end{align}
Such a power law behavior in the asymptotic form of the potentials is a characteristic of the background being extremal (as opposed to a non-extremal background for which the potential dies off exponentially as $u_*\to-\infty$). It is this power law behavior which gives rise to a branch cut  \cite{Ching:1995tj} (for small frequencies), extended in the negative imaginary axis, for the QNMs of $\Phi_{\pm}$. We will shortly see the appearance of this branch cut in our numerical plots of the QNMs.

\subsubsection {Stability} \label{stability}

Suppose $\Phi_{\pm}$ are QNMs with frequencies $\wn^{\pm}_n$. So, $\Phi_{\pm}$ are ingoing at the horizon and vanish asymptotically. Following \cite{Horowitz:1999jd}, we will see shortly that, using the properties of the potentials $V_{\pm}$, these two boundary conditions  constraint $\wn^{\pm}_n$. Since $\Phi_{\pm}$ are infalling at the horizon, one has $e^{-i\sqrt{3}\wn^{\pm}_n \tau}\Phi_{\pm}\sim e^{-i\sqrt{3}\wn^{\pm}_n(\tau+u_*)}$ where $\tau=r_0 t/L^2$ is the dimensionless time coordinate and $u_*$ is the tortoise coordinate defined in \eqref{tortoise}.  Note that $\tau+u_*$ is the dimensionless ingoing Eddington coordinate. Define 
\begin{align}\label{defpsi}
\Phi_{\pm}(u_*)=e^{-i\sqrt{3}\wn^{\pm}_n u_*}\Psi_{\pm}(u_*).
\end{align}
Substituting \eqref{defpsi} into \eqref{Scheq}, and writing the result in terms of the coordinate $u$, we obtain  
\begin{align}\label{eqpsi}
\left[u^2f(u)\Psi^{\prime}_{\pm}(u)\right]^{\prime}-2i\sqrt{3}\wn^{\pm}_n\Psi^{\prime}_{\pm}(u)-U_{\pm}(u)\Psi_{\pm}(u)=0,
\end{align}
where the potentials $U_{\pm}(u)$ are given by 
\begin{align}
U_{\pm}(u)=\frac{6}{u^3}\left(1\pm\sqrt{1+\qn^2}\right)+\frac{3}{u^2}\qn^2-uf'(u).
\end{align}
We multiply \eqref{eqpsi} by ${\bar\Psi}_{\pm}(u)$ and integrate the result to obtain 
\begin{align}\label{eqpsi1}
\left[u^2f(u){\bar\Psi}_{\pm}(u)\Psi^{\prime}_{\pm}(u)\right] ^{\infty}_{1}-\int_1^\infty du\left[u^2 f(u)\left|\Psi^{\prime}_{\pm}(u)\right|^2+2i\sqrt{3}\wn^{\pm}_n{\bar\Psi}_{\pm}(u)\Psi^{\prime}_{\pm}(u)+U_{\pm}(u)\left|\Psi_{\pm}(u)\right|^2\right]=0
\end{align}
where we have also performed an integration by parts. Using \eqref{defpsi} and \eqref{asymPhipm0} and the fact that $\Phi_{\pm}(u)$ are assumed to be QNMs, one obtains that as $u\to \infty$, $u^2\Psi^{\prime}_{\pm}(u)$ is finite and ${\bar\Psi}_{\pm}(u)$ vanishes. At the horizon,  both $\bar\Psi_{\pm}(u)$ and $\Psi^{\prime}_{\pm}(u)$ are finite and $f(u)$ vanishes. Thus, all together, one concludes that $\left[u^2f(u){\bar\Psi}_{\pm}(u)\Psi^{\prime}_{\pm}(u)\right] ^{\infty}_{1}=0$. Take the complex conjugate of \eqref{eqpsi1} and subtract the result from \eqref{eqpsi1} to obtain
\begin{align}\label{eqpsibarpsi}
\int_1^\infty du \left[\wn^{\pm}_n{\bar\Psi}_{\pm}(u)\Psi^{\prime}_{\pm}(u)+\bar\wn^{\pm}_n{\Psi}_{\pm}(u)\bar\Psi^{\prime}_{\pm}(u)\right]=0.
\end{align}
Integrating by parts the second term in \eqref{eqpsibarpsi}, and noting that ${\Psi}_{\pm}(u)$ vanishes at $u=\infty$, we obtain
\begin{align}\label{eqpsibarpasi1}
 \left({\rm Im}~\wn^{\pm}_n\right)\int_1^\infty du {\bar\Psi}_{\pm}(u)\Psi^{\prime}_{\pm}(u)=-\frac{i}{2}\bar\wn^{\pm}_n \left|\Psi_{\pm}(u=1)\right|^2.
\end{align}
Since  $\Psi_{\pm}(u=1)$ is finite, if  ${\rm Re}~\wn^{\pm}_n\neq 0$ equation \eqref{eqpsibarpasi1} then implies that ${\rm Im}~\wn^{\pm}_n\neq 0$. Substituting \eqref{eqpsibarpasi1} into \eqref{eqpsi1}, we obtain
\begin{align}\label{psifinaleq}
\int_1^\infty du\left[u^2 f(u)\left|\Psi^{\prime}_{\pm}(u)\right|^2+U_{\pm}(u)\left|\Psi_{\pm}(u)\right|^2\right]+\sqrt{3}\left({\rm Im}~\wn^{\pm}_n\right)^{-1}\left|\wn^{\pm}_n\right|^2\left|\Psi_{\pm}(u=1)\right|^2=0.
\end{align}
For all (real) values of $\qn$, $U_{+}(u)$ is always positive definite. This is because in order  for $U_{+}(u)$ to be always positive definite, one has to have  $3\qn^2+2\sqrt{1+\qn^2}-2\geq 0$, which is satisfied for all (real) values of $\qn$. Thus, the integral (with  subscript $+$) in \eqref{psifinaleq} never becomes negative. From equation \eqref{psifinaleq}, we then reach the conclusion that ${\rm Im}~\wn^{+}_n<0$. This result indicates that the $\Phi_{+}(u)$ fluctuation does not cause instability, as $e^{-i\sqrt{3}\wn^{+}_n \tau}\Phi_{+}$ decays in time for all $\wn^{+}_n$. For $U_{-}(u)$, on the other hand, it can easily be shown that the condition for it being positive definite turns out to be $3\qn^2-2\sqrt{1+\qn^2}-2\geq 0$ which is true only if $\qn^2\geq\frac{16}{9}$. Therefore, for $|\qn|\geq\frac{4}{3}$, the integral (with subscript $-$) in \eqref{psifinaleq}  is positive definite yielding ${\rm Im}~\wn^{-}_n(\qn)<0$. For $|\qn|<\frac{4}{3}$, there are ranges of $u$ for which $U_{-}(u)$ becomes negative. As a result, when $|\qn|<\frac{4}{3}$, without knowing the solution $\Psi_{-}(u)$, equation \eqref{psifinaleq} cannot by itself determine the sign of ${\rm Im}~\wn^{-}_n(\qn)$. However, as we will see later on, it is easy to numerically confirm  that even for $|\qn|<\frac{4}{3}$ one obtains ${\rm Im}~\wn^{-}_n(\qn)<0$, so the $\Phi_{-}(u)$ fluctuation does not cause instability either. 

In the background geometry we are considering, the quasinormal frequencies of $\Phi_{\pm}$ are further constrained.  Using equations \eqref{Scheq} together with the infalling boundary condition \eqref{defpsi}, one easily concludes that if $\wn^{\pm}_n(\qn)$ is a quasinormal frequency of $\Phi_{\pm}$, so is $-\wn^{\pm}_n(\qn)$, that is to say that the quasinormal frequencies appear as pairs with the same imaginary part but opposite real part. On the field theory side, the pairing of the poles of the retarded Green functions \eqref{Gytyt}--\eqref{Gyy} is due to the fact that the boundary theory is parity invariant.\footnote{This symmetry of the quasinormal modes would be broken, for example, in the presence of a magnetic field.}

\subsection{Numerical Analysis:  Matrix Method} 

To compute the quasinormal frequencies of $\Phi_{\pm}$ (in the extremal case) numerically, we use the ``matrix method'' of \cite{Leaver:1990zz}. Unlike the algorithm of \cite{Horowitz:1999jd} which is known not to work  for extremal black hole backgrounds\footnote{This is because for  extremal black holes, the horizon is an essential singularity.}, the matrix method of  \cite{Leaver:1990zz} is believed to be applicable here. The authors of \cite{Denef:2009yy} used this matrix method to find the quasinormal frequencies of a charged minimally-coupled scalar field in the extremal Reissner-Nordstr\"{o}m AdS$_4$ background. In what follows we first present our numerical results for the quasinormal frequencies of $\Phi_{\pm}$ in the extremal case, then compare their general features with the analogous quasinormal frequencies when the  Reissner-Nordstr\"{o}m  AdS$_4$ black hole background is non-extremal.  

\subsubsection{Zero Temperature}

For numerical purposes it is more convenient to switch to a new radial coordinate $z=1/u$. In this coordinate,  $z=1$ represents the horizon while $z=0$ is the asymptotic boundary.  We write the equations \eqref{phipmeq} in terms of the new coordinate $z$ and arrive at
\begin{align}\label{phipmeqz}
z^2\left[f(z)\Phi^{\prime}_{\pm}(z)\right]^{\prime}+\left[-zf'(z)+\frac{3z^2}{f(z)}\left(\wn^2-f(z)\qn^2\right)-6z^3\left(1\pm\sqrt{1+\qn^2}\right)\right]\Phi_{\pm}(z)&=0.
\end{align}
To apply the matrix method of \cite{Leaver:1990zz}, we first isolate the leading  behaviors of $\Phi_{\pm}(z)$ at the horizon and the boundary. The rest is then approximated by a power series in $z$ around a point $z_0$ such that the radius of convergence of the series covers both the horizon and the boundary. That point in our case is taken to be $z_0=\frac{1}{2}$.  Thus, we write
\begin{align}\label{seriesPhipm}
\Phi_{\pm}(z)=e^{i\frac{\wn}{\sqrt{12}(1-z)}}f(z)^{-i\frac{\sqrt{3}\wn}{9}} z \sum_{m=0}^M a^{\pm}_m(\wn,\qn) \left(z-\frac{1}{2}\right)^m.
\end{align}
Substituting \eqref{seriesPhipm} into \eqref{phipmeqz}, we obtain a set of $M+1$ linear equations for $M+1$ unknowns $a_p(\wn,\qn)$'s, and write them in a matrix form 
\begin{align}
\sum_{m=0}^M A^{\pm}_{mp}(\wn,\qn) a^{\pm}_p(\wn,\qn)=0.
\end{align}
Suppose now we choose a specific value for the spatial momentum $\qn$. The quasinormal frequencies $\wn^{\pm}_n$  for that particular value of $\qn$ are solutions to the following equation \begin{align}\label{det}
\det~A^{\pm}_{mp}(\wn^{\pm}_n,\qn)=0.
\end{align}
Since $\Phi_{\pm}$ in \eqref{seriesPhipm}  are  approximated by power series, the more terms kept in the series, the more accurate results one obtains for the quasinormal frequencies of $\Phi_{\pm}$. 
\begin{figure}[h]
$\begin{array}{c@{\hspace{0.07in}}c}
~~~~~(a)&~~~~~~~~~~~~~~(b)\\ \\
\includegraphics[width=3.1in]{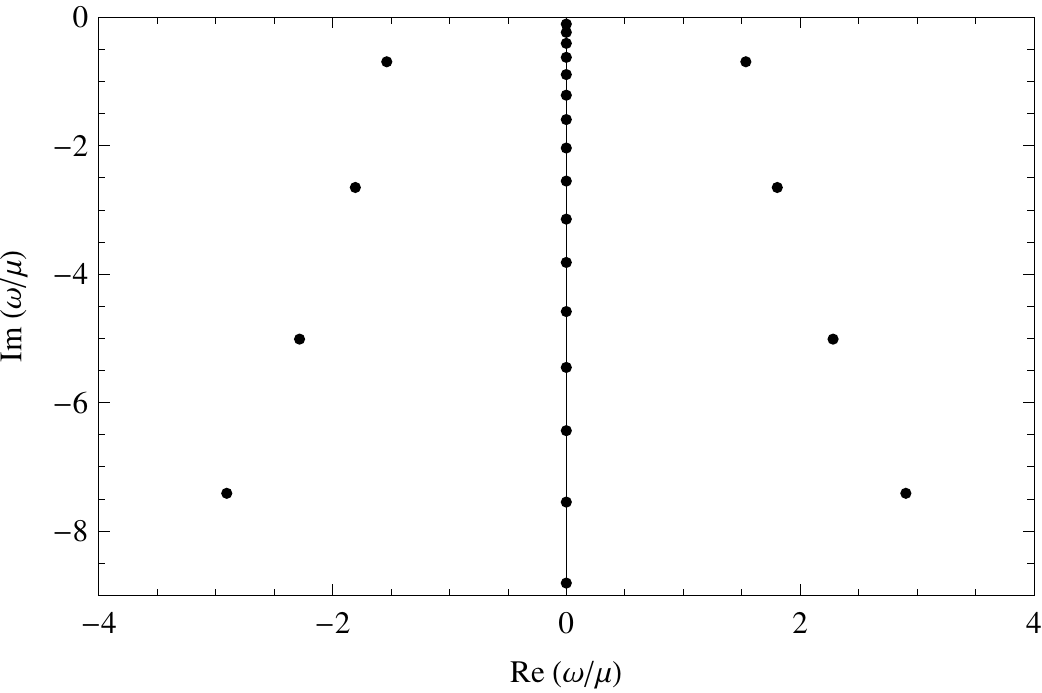}&\qquad
\includegraphics[width=3.1in]{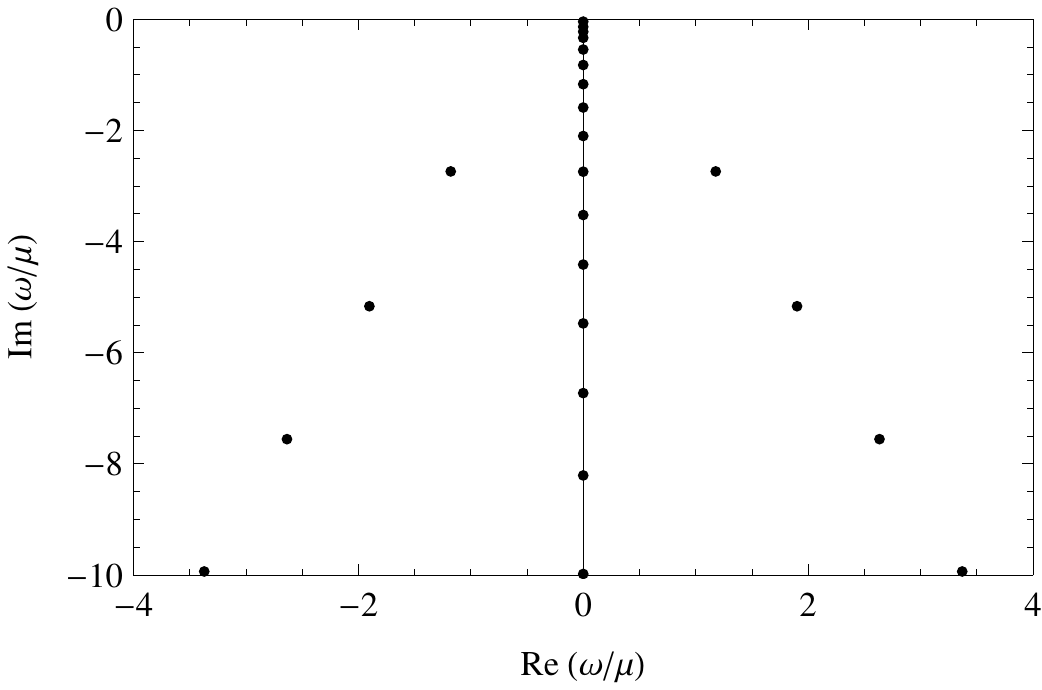}
\end{array}$
\caption[FIG. \arabic{figure}.]{\footnotesize{Electromagnetic and gravitational quasinormal frequencies (in the shear channel) of extremal Reissner-Nordstr\"{o}m  AdS$_4$ black hole.  Plot (a) shows the quasinormal frequencies of $\Phi_{+}$ and plot (b)  shows the quasinormal frequencies of $\Phi_{-}$.} For both plots, $\qn=1$ and $M=100$. As $M$ is increased the poles on the negative imaginary axis get closer to one another and form a branch cut in the $M\to\infty$ limit.}
   \label{figphipmzerotemp}
\end{figure}

\begin{figure}[h]
$\begin{array}{c@{\hspace{0.07in}}c}
~~~~~(a)&~~~~~~~~~~~~(b)\\ \\
\includegraphics[width=3.1in]{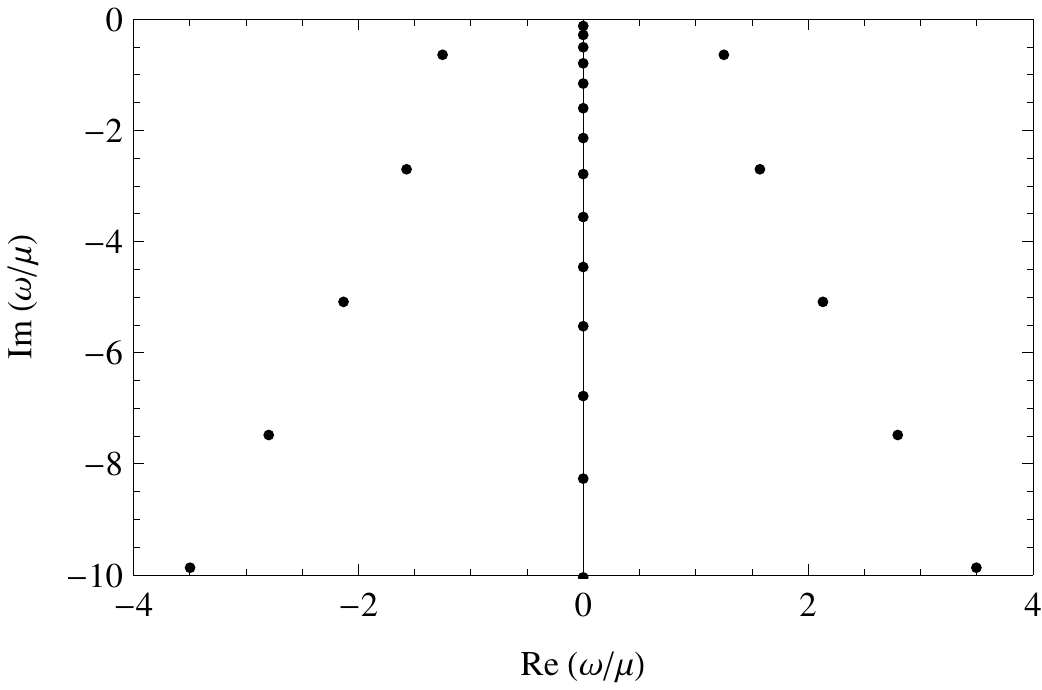}&\qquad
\includegraphics[width=3.1in]{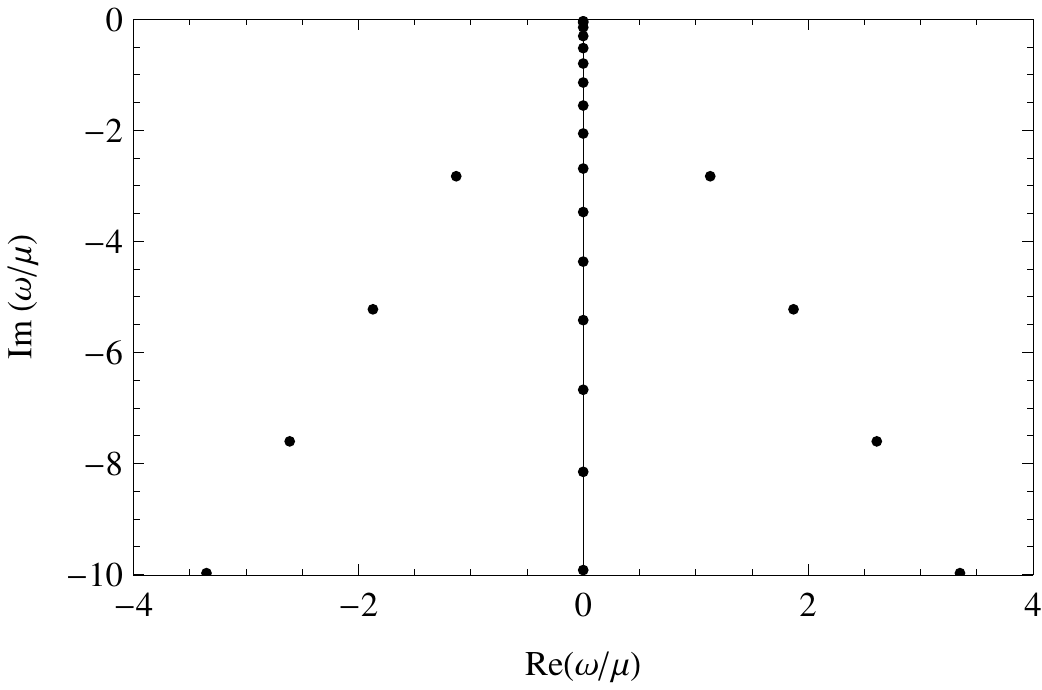}
\end{array}$
\caption[FIG. \arabic{figure}.]{\footnotesize{Quasinormal frequencies of $\Phi_\pm$ (in the extremal case) for $\qn=1/\sqrt{3}$. (a) Quasinormal frequencies of $\Phi_{+}$.  (b)  Quasinormal frequencies of $\Phi_{-}$. $M=100$ for both plots.}}
   \label{figphipm2zerotemp}
\end{figure}

The plots (a) and (b) in Figure \ref{figphipmzerotemp} show respectively the quasinormal frequencies of $\Phi_{+}$ and $\Phi_{-}$ for $\qn=1$, where, due to space limitations, we have only shown a handful of them. We used $M=100$ for both plots.  It can easily be verified numerically that, for both plots, as $M$ is increased the poles on the negative imaginary axis become closer to one another. This observation suggests that the sequence of discrete poles on the negative imaginary axis is due to approximating $\Phi_{\pm}$ in \eqref{seriesPhipm} by power series. In particular, although not numerically accessible, it should be the case that  in the $M\to\infty$ limit there is actually a branch cut \cite{Denef:2009yy}. This picture is consistent with the two arguments we presented  earlier for the existence of the branch point at the origin in the extremal case. Notice that the plots are symmetric under $\wn^{\pm}_n\to -\bar\wn^{\pm}_n$ as they should be. One also sees that the off-axis poles are approximately equally spaced and line up on almost straight lines.  Aside from the poles on the negative imaginary axis, which coalesce as $M$ is increased, the qualitative behavior of the plots does not change by increasing $M$. In particular, we do not see any poles  crossing into the real axis, indicating that these QNMs do not cause instabilities in the system. This observation is consistent with the argument we gave earlier in subsection \ref{stability}. The argument there was not powerful enough to determine the sign of ${\rm Im}~\wn^{-}_n(\qn)$ for $\qn\in(-\frac{4}{3},\frac{4}{3})$. The plot (b) of Figure \ref{figphipmzerotemp} (for which $\qn=1$) shows that the imaginary part of $\wn^{-}_n$ is negative. One can also plot the quasinormal frequencies of $\Phi_-$ for other values of $\qn$ in the range $(-\frac{4}{3},\frac{4}{3})$ and observe that ${\rm Im}~\wn^{-}_n$ is always negative. Indeed we have shown one such plot in Figure \ref{figphipm2zerotemp}(b).

The plots in Figure \ref{figphipm2zerotemp} show the quasinormal frequencies of $\Phi_{\pm}$ for  $\qn=1/\sqrt{3}$. Although not shown, we considered other values of $\qn$ and were able to verify that the qualitative patterns of the plots shown in Figures  \ref{figphipmzerotemp} and \ref{figphipm2zerotemp} stay the same as $\qn$ is varied. Excluding the poles on the negative imaginary axis, which form a branch cut in the  $M\to\infty$ limit, we observed that independent of what value $\qn$ takes, there are no small  quasinormal frequencies for $\Phi_{-}$, while there are only two for $\Phi_{+}$  (which are the ones closest to the real axis). Here, by small quasinormal frequencies we mean those which satisfy $|\wn^{\pm}_n|\leq1$. Therefore, one can conclude that at small frequencies $|\wn|\leq1$, there are only two single poles in the boundary theory retarded Green functions \eqref{Gytyt}--\eqref{Gyy}. Moreover,  those poles do \emph {not} approach zero as  $\qn\to 0$.

\begin{figure}[h]
$\begin{array}{c@{\hspace{0.07in}}c}
~~~~~~(a)&~~~~~~~~~~~~~(b)\\ \\
\includegraphics[width=3.1in]{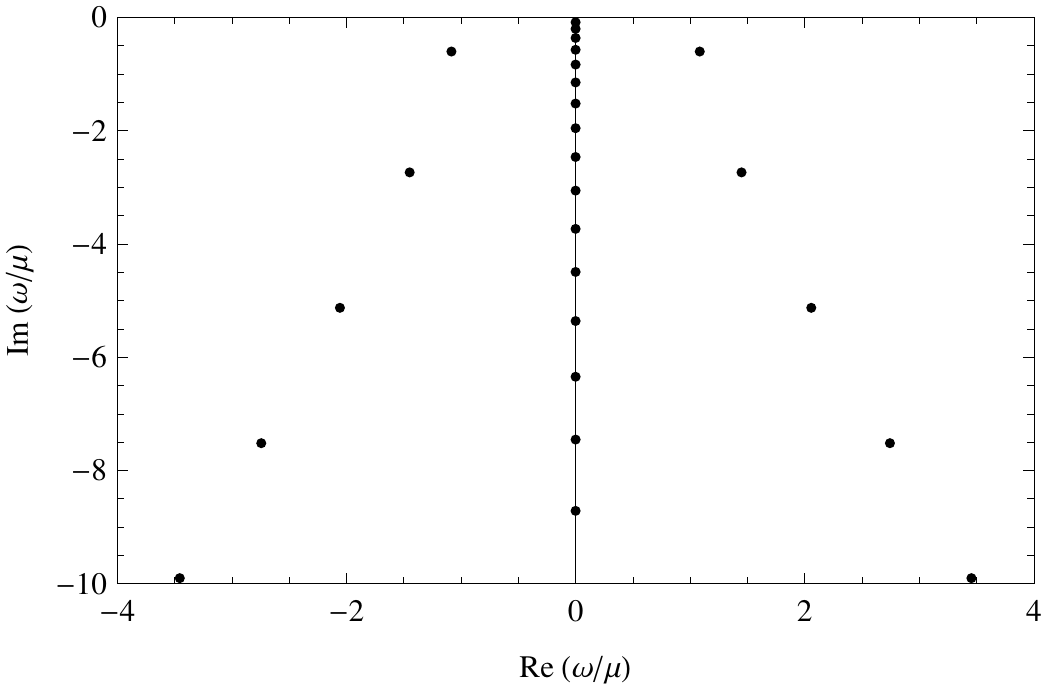}&\qquad
\includegraphics[width=3.1in]{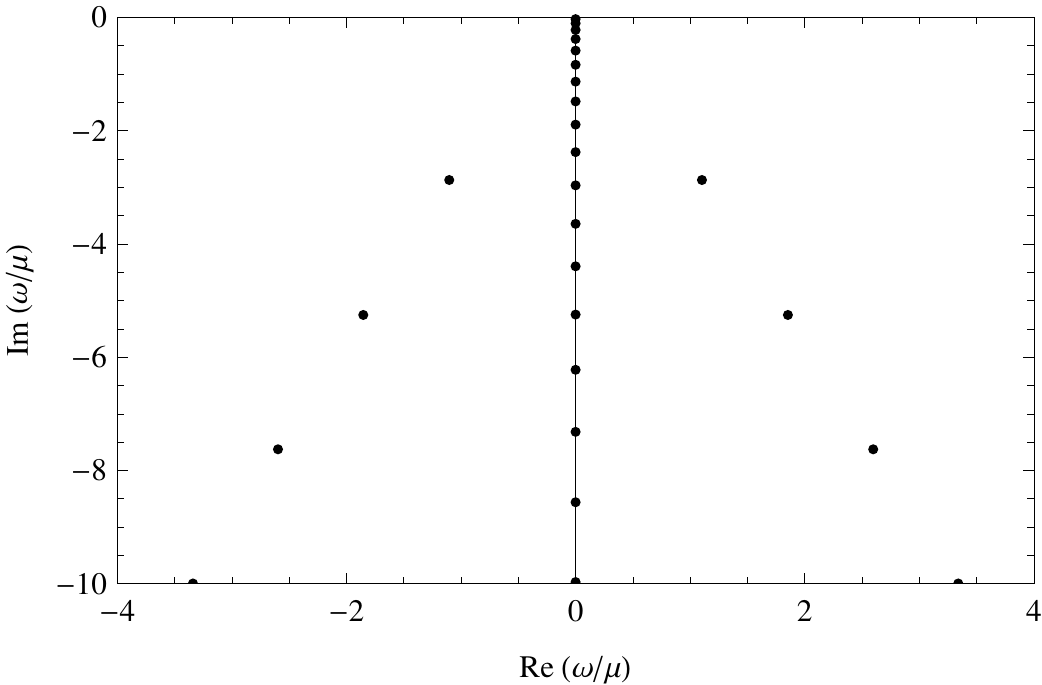}
\end{array}$
\caption[FIG. \arabic{figure}.]{\footnotesize{Quasinormal frequencies for $\qn=0$ of $a_y$ and ${h^x}_y$ in the extremal Reissner-Nordstr\"{o}m  AdS$_4$ black hole background. Plot (a)  shows the quasinormal frequencies of $a_y$  and plot (b)  are the quasinormal frequencies of ${h^x}_y$.} We set $M=100$ for both plots.}
   \label{figvanishingqzerotemp}
\end{figure}

The analysis at $\qn=0$ is quite a bit simpler, but is qualitatively similar.
In this case,  the relevant equations are\begin{align}
&z^2f(z){h^x}_y''(z)+\left[z^2f^{\prime}(z)-2zf(z)\right]{h^x}_y'(z)+\frac{3\wn^2}{f(z)}z^2 {h^x}_y(z)=0,\label{hxyzcoor}\\
&f(z)a^{\prime\prime}_y(z)+ f^{\prime}(z)a^{\prime}_y(z)+ \left( \frac{3\wn^2}{f(z)}-12z^2\right)a_y(z)=0,\label{ayzcoor}
\end{align}
which are obtained from equations \eqref{hxyzeroq} and \eqref{ayzeroq} by changing the radial coordinate to $z=1/u$. (Note that ${h^x}_y(z)$ is gauge invariant at $\qn=0$, while  $a_y(z)$ is gauge invariant for all $\qn$.) We can find the quasinormal frequencies of  ${h^x}_y(z)$ and $a_y(z)$ similar to the steps we performed above to obtain the quasinormal frequencies of $\Phi_{\pm}$. The only difference is that the normalizable mode of ${h^x}_y(z)$ falls off asymptotically as $z^3$.  Therefore  in writing the analog of  \eqref{seriesPhipm} for ${h^x}_y(z)$ the factor of $z$ should be replaced by $z^3$. The results are shown in Figures \ref{figvanishingqzerotemp}(a) and \ref{figvanishingqzerotemp}(b). Note that equation \eqref{hxyzcoor} is the same as the equation of motion for a minimally-coupled chargeless massless scalar in the background of the extremal Reissner-Nordstr\"{o}m AdS$_4$ black hole. The quasinormal frequencies of this scalar have been numerically computed in \cite{Denef:2009yy} using the same matrix method and a (density) plot resembling the one in Figure \ref{figvanishingqzerotemp} (b) was generated there.

The pattern we observe in the two plots of Figure \ref{figvanishingqzerotemp} is very much similar to the ones in Figures \ref{figphipmzerotemp} and \ref{figphipm2zerotemp}. The plots were obtained for $M=100$, but the pattern stays almost  the same for higher values of $M$. One still observes that  the poles on the negative imaginary axis coalesce while the locations of the poles on each side of the imaginary axis do not change much.  Also, we neither find poles crossing into the upper half plane, nor do we find poles which vanish. As it is seen from  Figure  \ref{figvanishingqzerotemp}(b), there are no (diagonally-oriented) quasinormal frequencies for ${h^x}_y$ at small $|\wn|\leq1$ frequencies  while there are two such frequencies for $a_y$ in that range. The quasinormal frequencies of $a_y$ in Figure \ref{figvanishingqzerotemp}(a) can be used to estimate, along the lines of \cite{Denef:2009yy}, the UV behavior of a one-loop contribution to the free energy of the extremal background by the gauge field $a_y$. Knowing the one-loop corrections to the free energy enables one to compute a ``$1/N$ correction'' to charge conductivity.

\subsubsection{Finite Temperature} 

It is instructive to  compare our results for the quasinormal frequencies of $\Phi_{\pm}$ in the extremal case to the quasinormal frequencies of $\Phi_{\pm}$ in the background of non-extremal Reissner-Nordstr\"{o}m  AdS$_4$ black hole. Written in terms of the $z$-coordinate, equations \eqref{finitephipmeq} take the form 
\begin{align}\label{phipmeqzfinite}
z^2\left[f(z)\Phi^{\prime}_{\pm}(z)\right]^{\prime}+\left[-zf'(z)+Q^2\frac{z^2}{f(z)}\left(\wn^2-f(z)\qn^2\right)-2Q^2g_{\pm}(\qn)z^3\right]\Phi_{\pm}(z)=0,
\end{align}
where $Q^2<3$ in the non-extremal case. The definition of $\Phi_{\pm}$ at finite temperature is given in \eqref{finitephipmdef}. 
\begin{figure}[h]
$\begin{array}{c@{\hspace{0.07in}}c}
~~~~~(a)&~~~~~~~~~~~~(b)\\ \\
\includegraphics[width=3.1in]{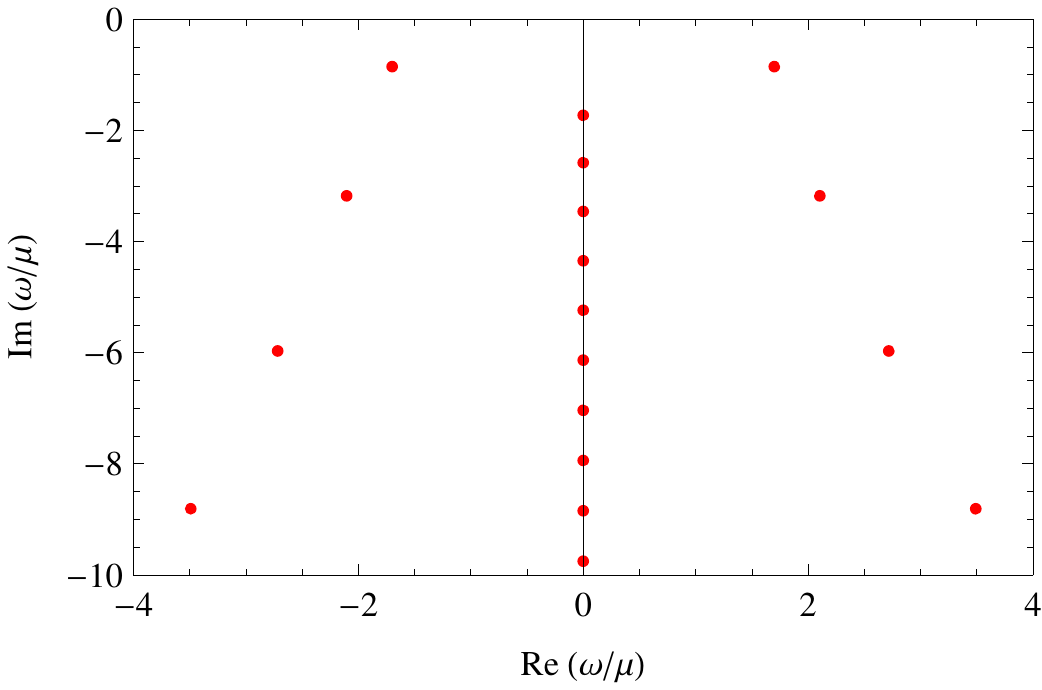}&\qquad
\includegraphics[width=3.1in]{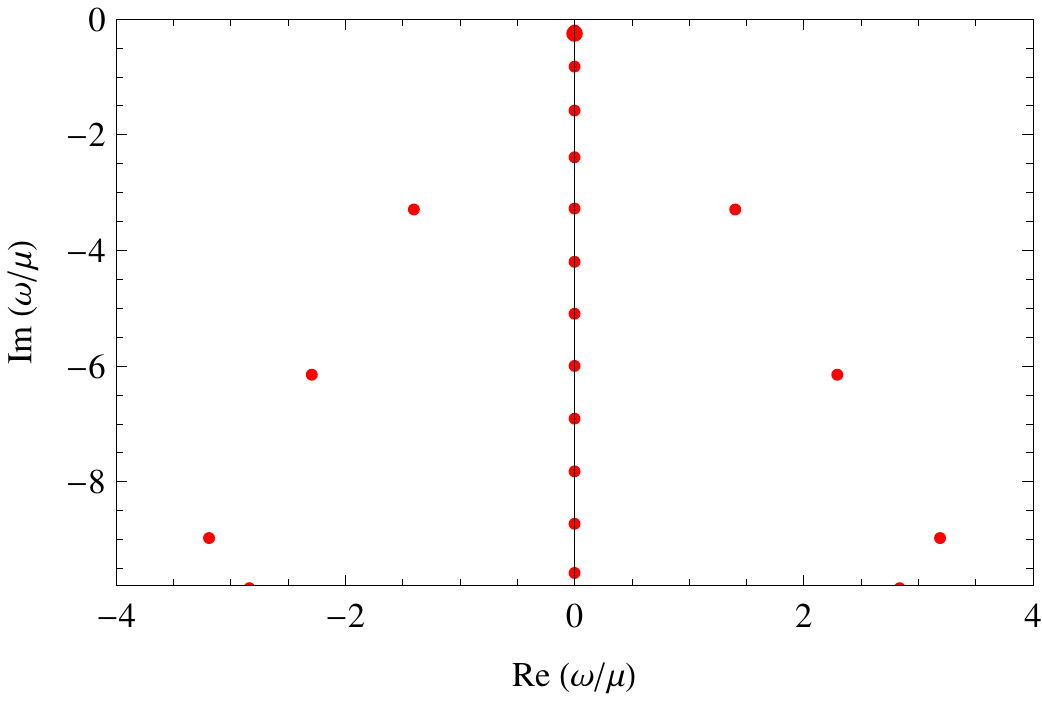}
\end{array}$
\caption[FIG. \arabic{figure}.]{\footnotesize{Quasinormal frequencies of $\Phi_{\pm}$ in non-extremal Reissner-Nordstr\"{o}m  AdS$_4$ black hole background. Plot (a) shows the quasinormal frequencies of $\Phi_{+}$ while plot (b)  shows  the quasinormal frequencies of $\Phi_{-}$.} For both plots, $\qn=1$,  $M=100$, and  $T=0.09\mu$. }
   \label{figphipmfinitetemp}
\end{figure}

In the non-extremal case, we also use  $\wn^{\pm}_n$ to  denote  the quasinormal frequencies of $\Phi_{\pm}$.  In order to find $\wn^{\pm}_n$ we use the matrix method of \cite{Leaver:1990zz}  and follow the exact same steps as we did in the extremal case except that \eqref{seriesPhipm} is now replaced by
\begin{align}\label{seriesPhipmfinite}
\Phi_{\pm}(z)=f(z)^{-i\wn/\tau} z \sum_{m=0}^M a^{\pm}_m(\wn,\qn) \left(z-\frac{1}{2}\right)^m,
\end{align}
where we have defined $\tau=4\pi T/\mu$, with $T$ and $\mu$ being the temperature and chemical potential, respectively. 
\begin{figure}[h]
\centerline{\includegraphics[width=3.6in]{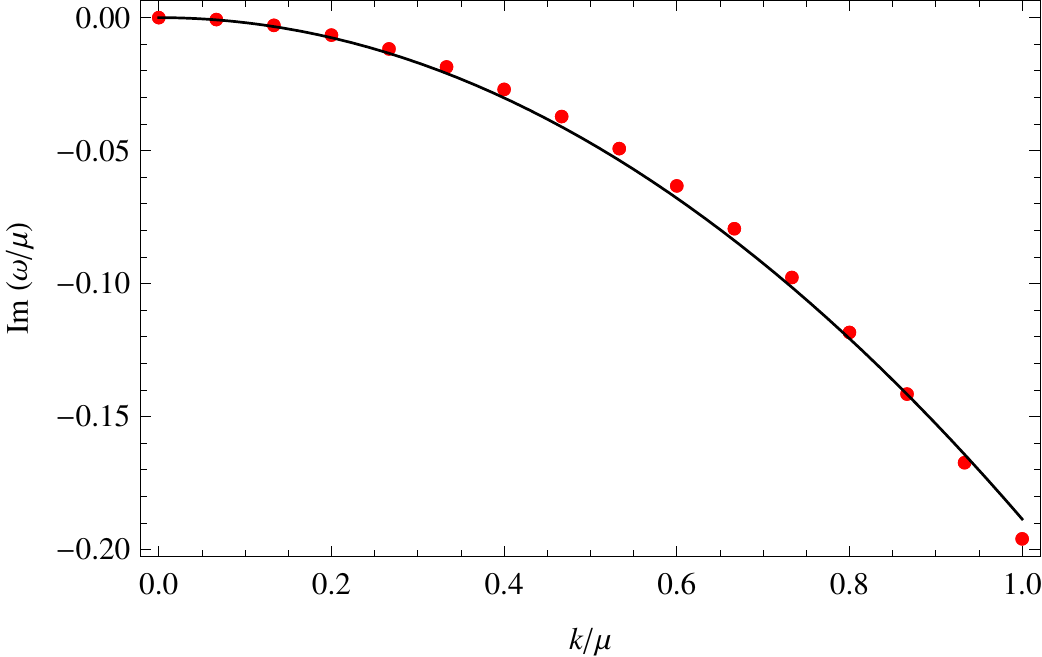}}
\caption[FIG. \arabic{figure}.]{\footnotesize{The lowest quasinormal frequency of $\Phi_{-}$ as function of $q$. The temperature is $T=0.09\mu$, and $M=100$. The solid black curve is the quadratic fit: $i \wn\simeq 0.17\qn^2$.}}
\label{figphipmfinitetempwvsq}
\end{figure}
The finite-temperature quasinormal frequencies of $\Phi_{\pm}$ have been plotted in   Figure \ref{figphipmfinitetemp}.  Only a handful of $\wn^{\pm}_n$ have been shown, though. To generate the plots, we chose  a temperature of $T=0.09\mu$. Also, the plots were obtained for $\qn=1$ and $M=100$. The branch cut in Figures \ref{figphipmzerotemp}(a) and \ref{figphipmzerotemp}(b) are replaced at finite temperature by a sequence of almost equally-distanced poles along the negative imaginary axis. Note that the poles are all in the lower half of the frequency plane, indicating that they do not give rise to instabilities.  As $\qn\to 0$, none of the quasinormal frequencies of $\Phi_{+}$ approach zero. On the other hand, the lowest quasinormal mode of  $\Phi_{-}$, shown in Figure \ref{figphipmfinitetemp}(b) by a larger red dot, approaches zero as  $\qn\to 0$. This is the leading hydrodynamic pole. Thus, one concludes that the finite temperature generalization of \eqref{Gytyt}--\eqref{Gyy} all have the same hydrodynamic pole, namely the diffusion constant $\cal D$ is the same for all of them.  In Figure \ref{figphipmfinitetempwvsq}, we plotted the lowest quasinormal frequency of $\Phi_{\pm}$ (shown in Figure \ref{figphipmfinitetemp}(b) by a larger red dot)  as a function of $\qn$, and fitted the plot by a quadratic function. The solid black curve on the plot shows the fit. From the slope of the fitted curve, we obtain ${\cal D}\simeq0.17/\mu$, with $T=0.09\mu$. Substituting the numerically obtained diffusion constant into the hydrodynamic equation $\eta=(Ts+\mu\rho){\cal D}$, where $\mu$, $s$ and $\rho$ are respectively given in \eqref{fmu} and \eqref{ece}, one easily obtains $\eta/s\simeq1/4\pi$.\footnote{The sign $\simeq$ is because of numerical uncertainties. One can of course show analytically that the ratio should be equal to $1/4\pi$ \cite{Iqbal:2008by, Benincasa:2006fu}.}
\begin{figure}[h]
$\begin{array}{c@{\hspace{0.07in}}c}
~~~~~~(a)&~~~~~~~~~~~~~(b)\\ \\
\includegraphics[width=3.1in]{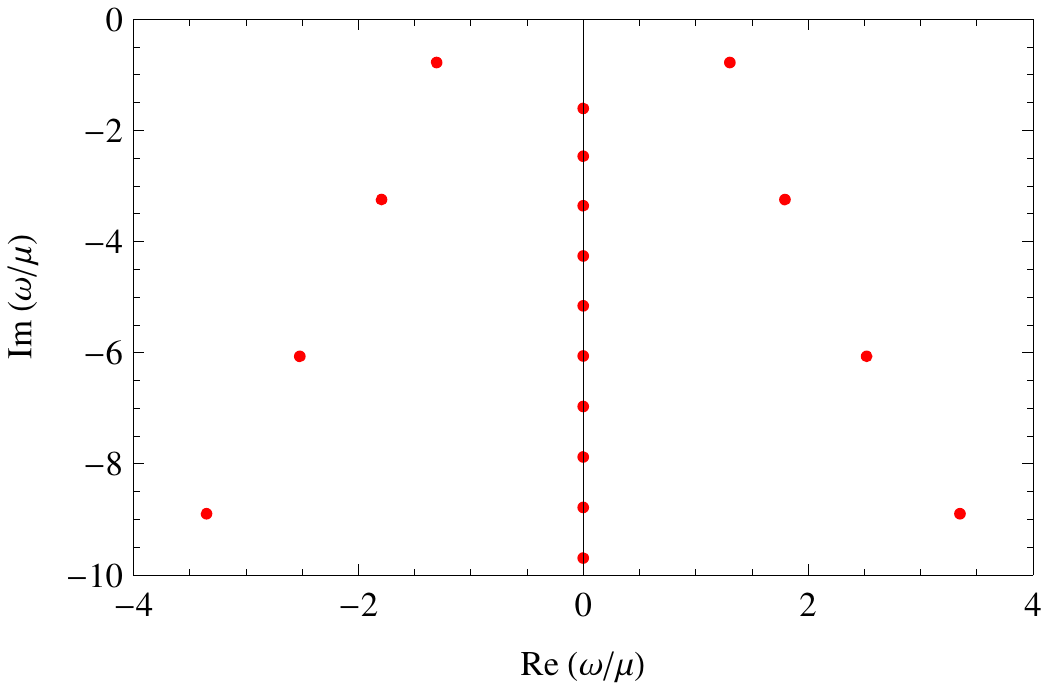}&\qquad
\includegraphics[width=3.1in]{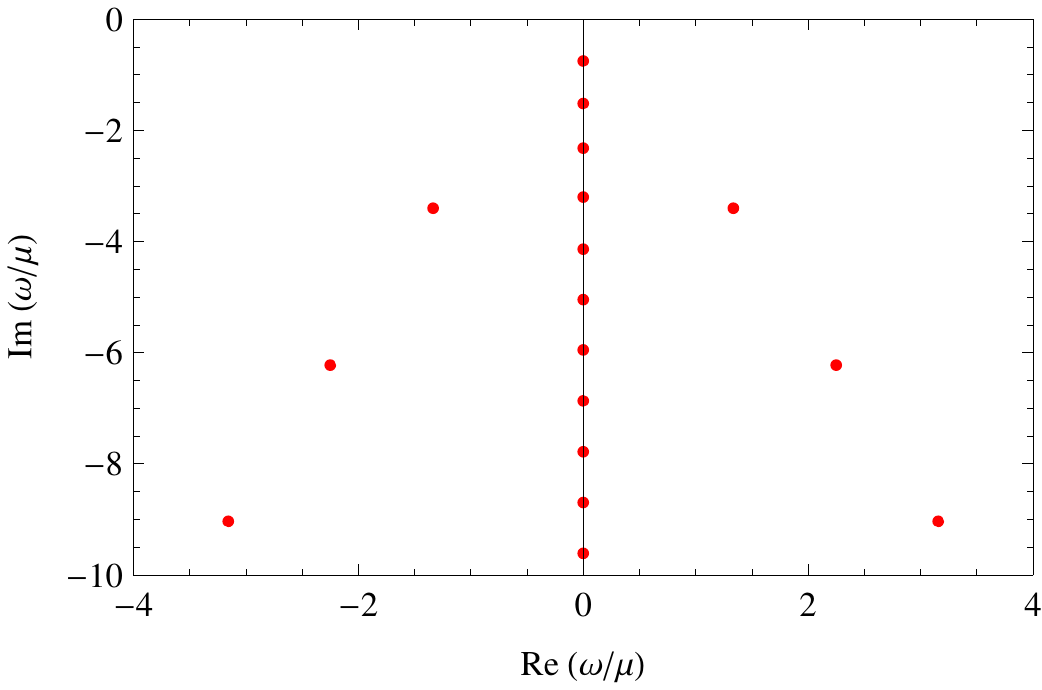}
\end{array}$
\caption[FIG. \arabic{figure}.]{\footnotesize{(a)  Finite-temperature quasinormal frequencies of $a_y$ at $\qn=0$. (b)  Finite-temperature quasinormal frequencies of ${h^x}_y$. The temperature is $T=0.09\mu$, and  $M=100$.}}
   \label{figvanishingqfinitetemp}
\end{figure}

We present the case of $\qn=0$ separately. The relevant equations in this case read
\begin{align}
&z^2f(z){h^x}_y''(z)+\left[z^2f^{\prime}(z)-2zf(z)\right]{h^x}_y'(z)+Q^2\frac{\wn^2}{f(z)}z^2 {h^x}_y(z)=0,\label{finitehxyzcoor}\\
&f(z)a^{\prime\prime}_y(z)+ f^{\prime}(z)a^{\prime}_y(z)+ Q^2\left( \frac{\wn^2}{f(z)}-4z^2\right)a_y(z)=0,\label{finiteayzcoor}
\end{align}
which are equations  \eqref{finiteeqhxyzeroq} and \eqref{finiteeqyzeroq} expressed in terms of the coordinate $z=1/u$. We find the finite-temperature quasinormal frequencies of  ${h^x}_y(z)$ and $a_y(z)$ following what we did  above to obtain the finite temperature quasinormal frequencies of $\Phi_{\pm}$. The only difference is that in writing the analog of  \eqref{seriesPhipmfinite} for ${h^x}_y(z)$ we replace the factor of $z$ by $z^3$. This is because asymptotically the normalizable mode of ${h^x}_y(z)$ falls off as $z^3$. 

The quasinormal frequencies of $a_y$ and ${h^x}_y$ have been plotted in Figure \ref{figvanishingqfinitetemp} for a temperature of $T=0.09\mu$ using $M=100$. We see that, as for $\qn\neq 0$, there is no branch cut at finite temperature: it is replaced by a sequence of almost equally-distanced poles along the negative imaginary axis. Also, as it is seen from the plots, the finite-temperature quasinormal frequencies of $a_y$ and ${h^x}_y$ stay away from the real axis.  To the extent that we have checked numerically, this behavior of the quasinormal frequencies is unchanged as $M$ is increased. 

\begin{figure}[h]
$\begin{array}{c@{\hspace{0.07in}}c}
~~~~~~(a)&~~~~~~~~~~~~~(b)\\ \\
\includegraphics[width=3.1in]{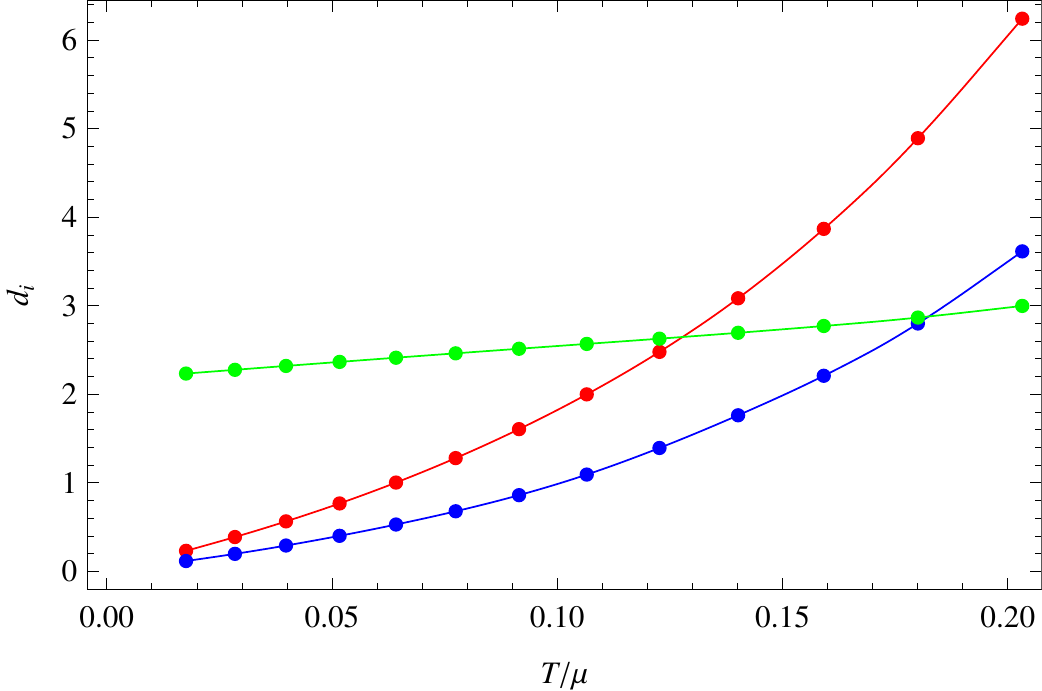}&\qquad
\includegraphics[width=3.1in]{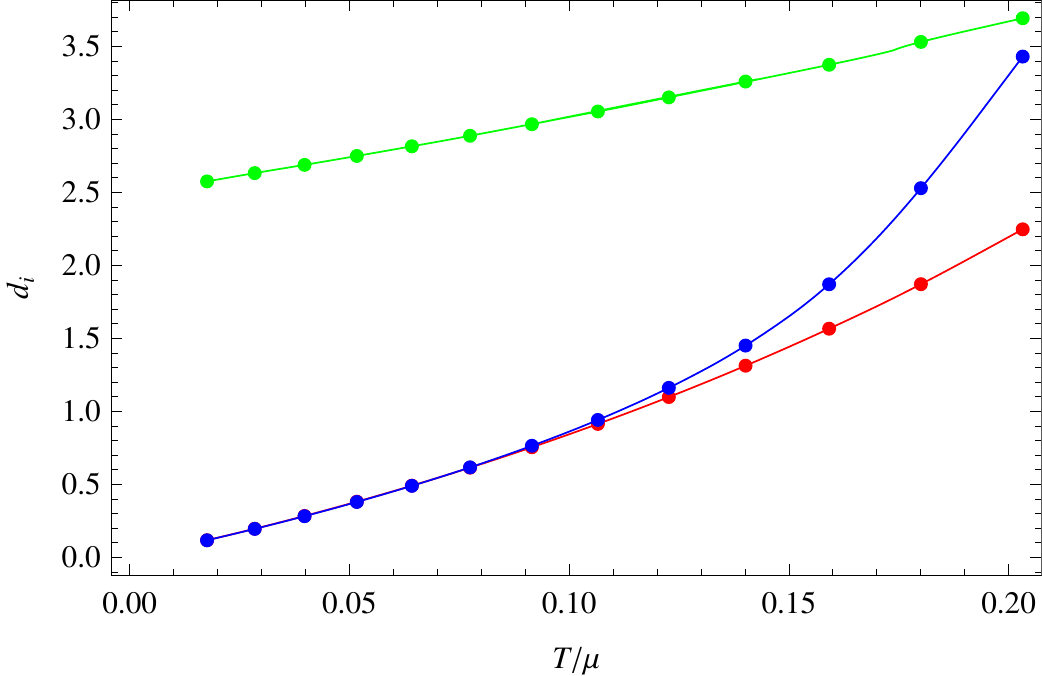}
\end{array}$
\caption[FIG. \arabic{figure}.]{\footnotesize{(a)  Plots of $d_1$ (red), $d_2$ (blue) and $d_3$   (green) as  function of $T/\mu$ for $a_y$. (b)  Plots of $d_1$, $d_2$ and $d_3$  as  function of $T/\mu$ for ${h^x}_y$. In the range of $T/\mu \in[0.017,0.091]$, the red and blue curves almost overlap. The plots have been generated with $M=150$.}}
  \label{distance}
\end{figure}

There does not appear to be a qualitative difference in the quasinormal spectrum between the $T=0$ results and the finite temperature spectrum for small temperatures. To see that more clearly, we consider the limit $T\to0$, holding $\mu$ fixed. For numerical simplicity, we have done this at $\qn=0$, but do not expect an appreciable difference at finite $\qn$. Referring to Figure \ref{figvanishingqfinitetemp}, we consider the following three quantities. We denote by $d_1$ the  difference between the smallest  quasinormal frequency on the negative imaginary axis and the origin, by $d_2$ the distance between the first two poles on the imaginary axis and by $d_3$, the distance between the two off-axis poles in the first quadrant closest to the origin. In Figure \ref{distance}(a), we demonstrate that as $T$ is lowered, holding $\mu$ fixed, both $d_1$ and $d_2$ approach zero, while $d_3$ remains finite. Approaching $T\to0$ requires ever-more numerical precision, but the plots suggest that $d_1$ and $d_2$ extrapolate to zero at $T=0$. This is consistent with the on-axis poles becoming the branch cut seen in the $T=0$ theory, as one might expect.

\section{Conclusions}

In this paper, we considered a boundary field theory dual to the extremal Reissner-Nordstr\"{o}m AdS$_{4}$ black hole, and analyzed the retarded two-point functions of the charge vector current and energy-momentum (tensor) operators at finite frequency and momentum.  The operators $\hat J_y$, $\hat T_{yt}$ and $\hat T_{xy}$ whose two-point functions we analyzed were in the shear channel of the boundary field theory, whose dual bulk modes $a_y$, ${h^y}_t$ and ${h^x}_y$ were odd under parity, $y\to -y$.  We found that  for generic momentum, the retarded correlators of the aforementioned vector and tensor operators exhibit emergent scaling behavior at low frequency. The operators $\hat J_y$, $\hat T_{yt}$ and $\hat T_{xy}$ gave rise in the IR to two sets of scalar operators ${\cal O}_{\pm}$ with conformal dimensions $\delta_{\pm}$. The low frequency scaling exponents of the spectral densities  in the boundary theory depended only on $\delta_{-}$ (or equivalently, $\nu_{-}$). We explored the analytic structure of the retarded correlators for finite momentum and small frequency, and argued that there should exist a branch cut in these correlators at the origin, as well as a series of metastable modes corresponding to isolated poles in the lower half of the complex frequency plane. Taking linear combinations of the dual bulk modes, we constructed the gauge invariant modes from which we were able to numerically compute the spectrum of the (shear-type) electromagnetic and gravitational quasinormal frequencies of the background. The numerical computations were in agreement with what we argued, analytically, for the existence of the branch cut in the retarded correlators as well as their poles being in the lower half of the complex frequency plane. Turning on temperature,  we numerically observed how the branch cut dissolves into a series of  poles on the negative imaginary axis, and determined the corresponding dispersion constant of the leading pole, in agreement with previous results. We performed the shear channel analyses in the $(2+1)$-dimensional boundary theory.  One can straightforwardly generalize the same type of analyses for higher dimensional boundary theories dual to extremal Reissner-Nordstrom AdS$_{d+1}$ black hole, and, not surprisingly,  reach the same conclusions as we did for the case of the $(2+1)$-dimensional boundary theory.

Although more involved, analyzing the retarded correlators of the operators in the sound channel of the $(2+1)$-dimensional boundary theory can similarly be performed. In the bulk, one deals with the modes which are even under the parity, $y\to -y$: $h_{tt}$, $h_{tx}$, $h_{xx}$, $h_{yy}$, $a_{t}$, $a_{x}$. Taking linear combinations of these modes, one can construct two gauge-invariant modes, and compute, either numerically or analytically for small $\wn$ and $\qn$, the spectrum of their quasinormal frequencies (having diagonalized the system of equations). The sound wave dispersion relation $\wn=\wn(\qn)$ can then be determined from the poles of, say, $G_{tt,tt} (\wn,\qn)$ at small $\wn$ and $\qn$. It is interesting to verify that $\wn(\qn)\to 0$ as $\qn\to 0$. The analysis for the sound channel of the $(2+1)$-dimensional boundary will be presented elsewhere.

\section*{Acknowledgments}

We would like to thank T. Andrade, S. Baharian, A. Buchel, E. Fradkin, S. Hartnoll, A. Hashimoto, C. Hoyos-Badajoz, A. Karch, J. McGreevy, A. Sinha and  D. Son for helpful discussions.  M.E. would like to especially thank S. Hartnoll for correspondence. R.G.L.\ and M.E.\ are supported by DOE grant FG02-91-ER40709, J.I.J. is supported by a Fulbright-CONICYT fellowship.

\appendix

\section{Finite Temperature Equations}
In the background of non-extremal Reissner-Nordstr\"{o}m  AdS$_4$ black hole, the linearized Einstein-Maxwell equations for the (odd-parity) modes ${h^y}_t$, ${h^x}_y$,  $a_{y}$ take the form  
\begin{align}
&f(r)\Big[r^4{h^y}_t''(r)+ 4r^3{h^y}_t'(r)+4Qr_0^2L^2a'_y(r)\Big]-\omega k L^4 {h^x}_y(r)-k^2L^4 {h^y}_t(r)=0, \label{finiteeqyt}\\ 
& f(r)\Big[r^4 f(r){h^x}_y''(r)+ \left[r^4f'(r)+4r^3f(r)\right]{h^x}_y'(r)\Big]+\omega^2 L^4 {h^x}_y(r)+\omega k L^4 {h^y}_t(r)=0,\label{finiteeqxy}\\ 
&f(r)\Big[r^4 f(r)L^2 a''_y(r)+ r^2L^2\left[r^2f'(r)+2rf(r)\right]a'_y(r)+Qr_0^2r^2{h^y}_t'(r)\Big]\nonumber\\
&+L^6\left[\omega^2-f(r)k^2\right]a_y(r) =0,\label{finiteeqy}
\end{align}
Also, the constraint equation (the $yr$-component of the linearized Einstein equations) reads
\bea\label{finitecons}
r^4 \omega~{h^y}_t'(r)+r^4 f(r) k~{h^x}_y'(r)+4Qr_0^2L^2 \omega~a_y(r)=0.
\eea
In the above expressions, $f(r)$ is given in (\ref{fmu}) and has a single zero at the (outer) horizon $r_0$. Note that $M=1+Q^2$. 

In the non-extremal case, we choose to rescale $\omega$ and $k$ by $\mu$. The dimensionless frequency and momentum, $\wn$ and $\qn$, are defined in \eqref{dimensionless}.  Also, like the extremal case, we work with the dimensionless radial coordinate $u=r/r_0$. The gauge-invariant combinations are given by the same expressions as in \eqref{ginvone} and \eqref{ginvtwo}:  $X(u)=\qn {h^y}_t(u) + \wn {h^x}_y(u)$ and $Y(u)= a_{y}(u)$.

At $\qn=0$, ${h^x}_y(u)$ becomes gauge-invariant. Also, the linearized Einstein-Maxwell equations \eqref{finiteeqyt}--\eqref{finitecons} give, in this case, two decoupled equations for ${h^x}_y(u)$ and $a_y(u)$
\begin{align}
&u^4 f(u){h^x}_y''(u)+ \left[u^4f'(u)+4u^3f(u)\right]{h^x}_y'(u)+Q^2\frac{\wn^2}{f(u)}{h^x}_y(u)=0,\label{finiteeqhxyzeroq}\\
&u^2 f(u)a''_y(u)+ \left[u^2f'(u)+2uf(u)\right]a'_y(u)+\frac{Q^2}{u^2}\left[\frac{\wn^2}{f(u)}-\frac{4}{u^2}\right]a_y(u)=0.\label{finiteeqyzeroq}
\end{align}
For finite (non-zero) $\qn$, on the other hand, it can easily be worked out that the linearized Einstein-Maxwell equations yield two coupled second-order differential equations for $X(u)$ and $Y(u)$. Similar to the extremal case studied in the bulk of the paper, these equations can be decoupled. We first define 
\begin{align}\label{finitephipmdef}
\Phi_{\pm}(u)&=-\mu \frac{\qn f(u)u^3}{\wn^2-f(u)\qn^2}X'(u)-\frac{2Q^2}{u}\left[\frac{2f(u)\qn^2}{\wn^2-f(u)\qn^2}+ug_{\pm}(\qn)\right]Y(u),
\end{align}
where
\begin{align}
g_{\pm}(q)=\frac{3}{4}\left(1+\frac{1}{Q^2}\right)\left(1\pm\sqrt{1+\frac{16}{9}\left(1+\frac{1}{Q^2}\right)^{-2}\qn^2}\right).
\end{align}
The decoupled equations then read 
\begin{align}\label{finitephipmeq}
\left[u^2f(u)\Phi^{\prime}_{\pm}(u)\right]^{\prime}+\left[uf'(u)+\frac{Q^2}{u^2f(u)}\left(\wn^2-f(u)\qn^2\right)-\frac{2Q^2}{u^3}g_{\pm}(\qn)\right]\Phi_{\pm}(u)&=0.
\end{align}


\end{document}